\documentclass[aps,prx,preprint,onecolumn,citeautoscript,nofootinbib,eqsecnum]{revtex4-2}
\usepackage[T1]{fontenc}
\usepackage{feynmp}
\usepackage{comment}
\usepackage{subcaption}
\usepackage{amsmath}
\usepackage{amssymb}
\usepackage{amsthm}
\usepackage{amsfonts}
\usepackage{bm,bbm}
\usepackage{braket}
\usepackage{xcolor}
\usepackage{pifont}
\usepackage{slashed}
\usepackage[mathscr]{euscript}
\usepackage[shortlabels]{enumitem}
\usepackage{tikz-feynman}
\usepackage[papersize={8.5in,11in}]{geometry}
\usepackage[colorlinks=true]{hyperref}
\hypersetup{
    bookmarks=true,         
    unicode=false,          
    pdftoolbar=true,        
    pdfmenubar=true,        
    pdffitwindow=false,     
    pdfstartview={FitH},    
    pdfsubject={},   
    pdfcreator={},   
    pdfproducer={}, 
    pdfkeywords={} {} {}, 
    pdfnewwindow=true,      
    colorlinks=true,       
    linkcolor=magenta, 
    citecolor=blue,        
    filecolor=magenta,      
    urlcolor=blue           
} 

\usepackage{amsfonts}
\usepackage{upgreek}
\usepackage{slashed}
\usepackage{latexsym}

\newcommand{\beq}{\begin{equation}}
\newcommand{\eeq}{\end{equation}}
\def\bea{\begin{eqnarray}}
\def\eea{\end{eqnarray}}

\renewcommand{\approx}{\simeq}

\renewcommand{\Im}{\text{Im}}

\definecolor{wrongultramarine}{rgb}{1,0.5,0}

\newcommand{\sgn}{{\rm sgn\,}}

\newcommand{\iw}{i\omega}
\newcommand{\om}{\omega}

\begin{document}


\title{Superconductivity of non-Fermi liquids\\ described by Sachdev-Ye-Kitaev models}

\author{Chenyuan Li}
\affiliation{Department of Physics, Harvard University, Cambridge MA 02138, USA}
\author{Subir Sachdev}
\affiliation{Department of Physics, Harvard University, Cambridge MA 02138, USA}
\author{Darshan G. Joshi}
\affiliation{Department of Physics, Harvard University, Cambridge MA 02138, USA}

\date{\today
\\
\vspace{0.4in}}

\begin{abstract}
We investigate models of electrons in the Sachdev-Ye-Kitaev class with random and all-to-all electron hopping, electron spin exchange, and Cooper-pair hopping. An attractive on-site interaction between electrons leads to superconductivity at low temperatures. Depending on the relative strengths of the hopping and spin exchange, the normal state at the critical temperature is either a Fermi-liquid or a non-Fermi liquid. We present a large-$M$ (where spin symmetry is enlarged to SU$(M)$) study of the normal state to superconductor phase transition. We describe the transition temperature, the superconducting order parameter, and the electron spectral functions. We contrast between Fermi liquid and non-Fermi liquid normal states: we find that for weaker attractive on-site interaction there is a relative enhancement of $T_c$ when the normal state is a non-Fermi liquid, and correspondingly a strong deviation from BCS limit. Also, the phase transition in this case becomes a first-order transition for strong non-Fermi liquids. On the other hand, for stronger on-site interaction, there is no appreciable difference in $T_c$ between whether the superconductivity emerges from a Fermi liquid or a non-Fermi liquid.  Notable features of superconductivity emerging from a non-Fermi liquid are that the superconducting electron spectral function is different from the Fermi-liquid case, with additional peaks at higher energies, and there is no Hebel-Slichter peak in the NMR relaxation rate in the non-Fermi liquid case.
\end{abstract}

\maketitle
\newpage
\section{Introduction}
\label{sec:intro}

The classic BCS theory provides a highly successful description of the onset of superconductivity (SC) from a Fermi liquid (FL). However, in modern correlated electron materials, the normal state at the onset of higher temperature superconductivity is usually not a Fermi liquid. Below the critical temperature, basic aspects of the BCS superconducting state (such as the breaking of U(1) gauge symmetry by an electron pair condensate) continue to hold, but numerous quantitative details on the critical temperature, superconducting gap amplitude, and electron spectral function are not described by BCS theory.

A popular class of theories for the onset of superconductivity from a non-Fermi liquid (NFL) focus on a normal state which has a Fermi surface coupled to a critical boson \cite{Chubukov21,Chubukov22,Berg20,Mross15,Mandal2016,Esterlis:2021eth}. The boson could represent a symmetry breaking order parameter at a quantum critical point, or an emergent excitation associated with spin liquid physics. This critical boson plays a dual role---it leads to the breakdown of quasiparticles in the normal state, and it also leads to superconductivity at low temperature ($T$) by inducing pairing  between the underlying electrons. The precise manner in which the non-Fermi liquid gives way to superconductivity at low $T$ is not well understood, and remains a topic of great interest.

In this paper, we will address the interplay between the non-Fermi liquid and superconductivity using a different class of simpler and more tractable models. These models do not have much spatial structure because of the presence of all-to-all hopping and interactions. However, they have the virtue of being exactly solvable, and so can describe the competition between the different energy scales in a quantitative manner. We consider the Sachdev-Ye-Kitaev (SYK) type of models \cite{SY92,kitaev2015talk}, which are a rare class of solvable models leading to non-Fermi liquid phases \cite{syk_review}. Models in this class have been recently studied in different contexts of strongly correlated systems. In this work we consider a model of electrons with an attractive on-site interaction. In the spirit of SYK models, we consider random and all-to-all hopping, exchange interaction, and Cooper-pair hopping. This model was previously considered by us and for weak interaction an anomalous metal phase (or a Bose metal) was shown to exist \cite{Li2021} in the proximity of superconducting phase. In this work our focus is on the superconducting phase, and the associated thermal phase transition. Depending on the relative strength of the hopping amplitude and exchange interaction, the normal state at higher temperatures is either a FL or a NFL. Thus our model allows us to systematically investigate the emergence of superconductivity by continuously tuning between FL and NFL normal states. 
Moreover, we show that SC emerging from a NFL has certain unique features in the spectral function that are absent in the case of a FL-SC transition. 

There have been previous studies of superconductivity in SYK models \cite{Klebanov2020, Hurtubise2021, Wang2020,Kamenev2020,Esterlis2019,Patel2018,Chowdhury2020,Chudnovskiy2022,Sahoo2020}. However, our model is distinct from the previously considered models. In our model in Eq. (\ref{eq:Ham}), we start with a SU($2$) spin symmetry (see $H_{J}$ in Eq. (\ref{eq:H_J})), just as in the original Sachdev-Ye (SY) model \cite{SY92}. In previous models the random and all-to-all SYK term is in general not SU$(2)$ symmetric:  
in Refs.~\cite{Klebanov2020, Hurtubise2021} a general Hamiltonian of two coupled SYK models is considered, which has a SU$(2)$ symmetry only at a special point ($\alpha=1/4$ in the notation used in Ref. \cite{Klebanov2020}), and it corresponds to the zero hopping limit with $U=t=L=0$ in our model. However, it is shown in Refs. \cite{Klebanov2020, Hurtubise2021} that at this SU$(2)$ symmetric point there is no superconductivity, which is consistent with our results.  Ref.~\cite{Kamenev2020} also examined models without any hopping, but did examine finite $N$ corrections. The models of Refs.~\cite{Patel2018,Chowdhury2020} are related to the one examined here, but
with lattice rather than random matrix hopping: the lattice dispersions and all-to-all random hopping for electrons lead to equations with similar solutions \cite{syk_review}. Because of the simpler form of our equations, we are able to present spectral functions within the superconducting phase across the full range of the crossover between the FL and NFL cases.

The plan of the paper is as follows. In Sec. \ref{sec:BCS} we first study SC in a simple model of attractive Hubbard model with random and all-to-all hopping. Then we introduce our model in Sec. \ref{sec:mod} and discuss the saddle-point equations. These equations are solved to obtain the normal state and SC solutions in Sec. \ref{sec:num}. Therein we discuss several observables. Finally we conclude in Sec. \ref{sec:dis}. Technical details are provided in Appendices. 


\section{Random matrix Bogoliubov-de Gennes theory}
\label{sec:BCS}

Before we dive into the actual model and its detailed analysis, let us first consider a simpler case. 
In this section we present a BCS theory of superconductivity for a Hubbard model with attractive on-site interaction $U$ along with a random and all-to-all hopping.
Our main purpose here is to introduce the formalism in a more familiar setting. Curiously, the spectral functions in the superconducting state in this simple model do not appear to have been obtained earlier, although there have been results for other quantities for finite $N$ \cite{Yuz03,Yuz05}.

We consider a model of electrons $c_{i\alpha}$, with $i=1 \ldots N$ a site index, and $\alpha = 1 \ldots M$ a USp($M$) index. We have thus enlarged the usual SU($2$) spin symmetry.
The USp($M$)
group, $M$ even, is defined by the set
of $M\times M$ unitary matrices $\mathcal{U}$
such that
\begin{equation}
\mathcal{U}^{T} \mathcal{J} \mathcal{U} = \mathcal{J} \,,
\label{dsp2}
\end{equation}
where
\begin{equation}
\mathcal{J}_{\alpha\beta} = \mathcal{J}^{\alpha\beta} = \left( \begin{array}{cccccc}
  & 1 & & & & \\
-1 &  & & & & \\
 & &  & 1  & & \\
 & & -1 &  & & \\
 & &  & & \ddots & \\
 & & & & & \ddots
\end{array} \right)
\end{equation}
is the generalization of the $\varepsilon$ tensor to $M>2$. It is
clear that USp$(M) \subset$ SU($M$) for $M>2$, while USp(2) $\cong$ SU(2). We will consider SYK-like models on $N$ sites with USp$(M)$ symmetry, and take the $N \rightarrow \infty$ limit followed by the $M \rightarrow \infty$ limit. We don't expect the large $M$ limit to significantly modify the results, as discussed in Ref.~\cite{syk_review}; the large $N$ limit is more significant, and there can additional phases at finite $N$, as discussed in Refs.~\cite{Kamenev2020,Chudnovskiy2022}.

We shall calculate the electron spectral density using a set of saddle-point equations, which we derive below. 
We consider an attractive Hubbard model on a random hopping matrix with the Hamiltonian,
\beq
\label{eq:bcs_ham}
H_{tU} = - \frac{1}{\sqrt{N}} \sum_{i<j} t_{ij} \left( c_{i \alpha}^\dagger c_{j}^{\alpha} + c_{j \alpha}^\dagger c_{i}^ {\alpha} \right) + \sum_i \left[
- \mu c_{i \alpha}^\dagger c_{i}^{\alpha} + \frac{U}{2M} \left| \mathcal{J}^{\alpha\beta} c_{i \alpha}^\dagger c_{i \beta}^{\dagger} \right|^2  \right] \,,
\eeq
where $t_{ij}$ is a {\it real} random number with zero mean and root-mean-square value $t$, $N$ is the number of sites, $\mu$ is the chemical potential and $U<0$ is the attractive on-site interaction. In terms of the electron annihilation (creation) operator, $c_{\alpha} (c_{\alpha}^\dagger)$, the number operator $n_{\alpha} = c_{\alpha}^\dagger c_{\alpha}$. 

We perform a disorder average to obtain the following action:
\bea
\overline{\mathcal{S}} &=& \sum_i \int d \tau \left[ c_{i \alpha}^\dagger (\tau) \left( \frac{\partial }{\partial \tau} - \mu \right) c_{i}^ {\alpha} (\tau) + \frac{U}{2M} \left| 
\mathcal{J}^{\alpha\beta} c_{i \alpha}^\dagger (\tau) c_{i \beta}^{\dagger} (\tau) \right|^2 \right] \nonumber \\
&~&~~~
+ \frac{t^2}{2 N} \int d \tau d \tau' \left[ \left| \sum_i c_{i \alpha}^\dagger (\tau) c_{i}^{ \beta} (\tau') \right|^2 - \left| \sum_i c_{i \alpha}^\dagger (\tau) c_{i \beta}^\dagger (\tau') \right|^2 \right] \,, \label{Sbar}
\eea
where $\tau$ is the imaginary time. 
Note that we have ignored here the replica indices as they are not significant for the present discussion. 
Next, we proceed by the $G$-$\Sigma$ method used for SYK models. We introduce the normal and anomalous Green's functions $G$ and $F$ respectively, as well as the normal and anomalous self energies $\Sigma$ and $\Phi$ respectively. We can then write the path integral as
\beq
\mathcal{Z}_{tU} = \int \mathcal{D} G \mathcal{D} F \mathcal{D} \Sigma \mathcal{D} \Phi \mathcal{D} c \, \exp( - \mathcal{S}_0 - \mathcal{S}_1)\,,
\eeq
where, initially, the role of the self energies is to impose delta functions which define the Green's functions as two-point fermion correlators. Let us now look at the two contributions in the action. First we have,
\bea
\mathcal{S}_0 &=& \int d \tau   \sum_i c_{i \alpha}^\dagger (\tau) \left( \frac{\partial }{\partial \tau} - \mu \right) c_{i}^ {\alpha} (\tau)  + \int d\tau d\tau' \Sigma (\tau, \tau') \left[ \sum_i c_{i \alpha}^\dagger (\tau) c_i ^{ \alpha} (\tau')
- N M G(\tau' , \tau) \right] \nonumber \\
&~&~~~+ \int d \tau d \tau' \frac{\Phi (\tau , \tau')}{2} \left[ 
 \mathcal{J}^{\alpha\beta} \sum_i c_{i\alpha}^\dagger (\tau) c_{i\beta}^{\dagger} (\tau') + NM F^\ast (\tau, \tau') \right] \nonumber \\
&~&~~~ - \int d \tau d \tau' \frac{\Phi^\ast (\tau , \tau')}{2} \left[  \mathcal{J}_{\alpha\beta} \sum_i c^{i \alpha} (\tau) c_i^{ \beta} (\tau') - NM F (\tau, \tau') \right] \,. 
\label{S0}
\eea
For the interaction terms in (\ref{Sbar}), we need to introduce additional Hubbard-Stratonovich terms which decouple the quartic fermion interactions, and then use the large $M$ limit to replace these fields by their saddle-point values. This procedure has been carried out explicitly for a related model in Ref.~\cite{Christos:2021wno}, and we do not display the intermediate steps here. Assuming the saddle-point has USp($M$) symmetry, we can obtain the final answer more directly simply by the following identifications in the interaction terms:
\bea
c_{ \alpha}^\dagger (\tau) c^{ \beta} (\tau') & \Rightarrow & \delta_\alpha^\beta \, G(\tau', \tau) \,, \nonumber \\ 
 c^{\alpha} (\tau) c^{ \beta} (\tau') & \Rightarrow & - \mathcal{J}^{\alpha\beta} F(\tau, \tau')
 \label{Me1}
\eea
In this manner we obtain the second contribution in the action,
\beq
\frac{\mathcal{S}_1}{NM} = 
 \frac{U}{2} \int d \tau \left| 
F(\tau,\tau) \right|^2 + \frac{t^2}{2} \int d \tau d \tau' \left[ G(\tau, \tau') G(\tau', \tau) - F(\tau, \tau') F^\ast (\tau', \tau) \right] \,.
\label{S1}
\eeq

Now we take the variational derivative of the action with respect to $G$ and $F^\ast$, and obtain the saddle-point equations,
\bea
\Sigma (\tau, \tau') &=& t^2 G (\tau, \tau') \,, \nonumber \\
\Phi (\tau, \tau') &=& - U F(\tau, \tau) \delta(\tau - \tau') + t^2 F(\tau, \tau') \,.
\label{saddle1}
\eea
These equations have to be supplemented by the Dyson equations obtained from the single-site action for the fermions, which follows from the first 2 spin components of the action $\mathcal{S}_0$,
\beq
\mathcal{S}_c = T \sum_{\omega} \left( c_{\uparrow}^\dagger (i\omega),  c_{\downarrow} (-i\omega) \right)\left(
\begin{array}{cc} -i \omega - \mu + \Sigma (i\omega) & \Phi (i \omega) \\
\Phi^\ast(i \omega) & -i \omega + \mu - \Sigma (-i\omega) \end{array}
\right)
\left( \begin{array}{c} c_{\uparrow} (i\omega) \\ c_{\downarrow}^\dagger (-i \omega)
\end{array} \right) 
\,,
\eeq
where $T$ is the temperature.
We can now write down the combined saddle point equations,
\bea
G_\Sigma (i\omega) &\equiv & \frac{1}{i \omega +\mu - \Sigma (i\omega)} \,, \nonumber \\
\Sigma (i\omega) &=& t^2 G(i \omega) =  t^2 \frac{\left[G_\Sigma (-i\omega) \right]^{-1}}{|\Phi(i\omega)|^2 + \left[G_\Sigma (i\omega) G_\Sigma (-i\omega)\right]^{-1}} \,, \nonumber \\
\Delta &=& -U T \sum_\omega \frac{\Phi(i\omega)}{|\Phi(i\omega)|^2 + \left[G_\Sigma (i\omega) G_\Sigma (-i\omega)\right]^{-1}} \,, \nonumber \\
F(i\omega) &=& \frac{\Phi(i\omega)}{|\Phi(i\omega)|^2 + \left[G_\Sigma (i\omega) G_\Sigma (-i\omega)\right]^{-1}} \,, \nonumber \\
\Phi (i \omega) &=& \Delta + t^2 F(i \omega) \,.
\label{eall}
\eea
The normal and anomalous Green's function in the superconducting state are $G(i \omega)$ and $F(i \omega)$ along the Matsubara frequency axis, while $G_\Sigma (i \omega)$ is an intermediate quantity defined for notational convenience; $G(i \omega) = G_\Sigma (i \omega)$ only in the normal state where $\Delta=F(i \omega)=0$.

It is useful to first solve these equations in the normal state solution by setting $\Delta = F(i \omega) = 0$, which yields for $\mu < 2t$
\beq
G(i \omega) \equiv G_0 (i \omega) = \frac{i \omega +\mu}{2 t^2}  - i \frac{\mbox{sgn}(\omega)}{2t^2} \sqrt{4t^2 + (\omega - i\mu)^2}  \,, \label{Gnormal}
\eeq
where the sign in front of the square-root is discontinuous across the real frequency axis, and is chosen so that $G_0 (z) \sim 1/z$ as $|z| \rightarrow \infty$.
This yields the expected semi-circle density of states.

Next, we can linearize the equations (\ref{eall}) in $\Delta$ at $T>0$, and so obtain the superconducting critical temperature $T_c$. We find the condition
\beq
1 = -U T \sum_{\omega_n} \frac{G_0 (i\omega_n) G_0 (-i \omega_n)}{1 - t^2 G_0 (i \omega_n) G_0 ( - i\omega_n)} \,, \label{eqTc}
\eeq
with $\omega_n$ a Matsubara frequency. At small $|\omega_n|$ we obtain from Eq. (\ref{Gnormal}) that
\beq
t^2 G_0 (i \omega_n) G_0 (- i \omega_n) = 1 - \frac{2 |\omega_n|}{\sqrt{4t^2 - \mu^2}} + \mathcal{O}(\omega_n^2)\,.
\eeq
We can now observe that the denominator in Eq. (\ref{eqTc}) has a singularity at $\omega_n = 0$, which yields the BCS log divergence. This implies that there is superconductivity at $T=0$ for infinitesimal negative $U$.

We can analytically solve Eqs. (\ref{eall}) at $T=0$ to linear order in $\Delta$ for general $\mu$. Such a solution will be valid for $|\omega|, \sqrt{4t^2 - \mu^2} \gg \Delta$. We find,
\bea
F(i \omega) &=& \Delta\frac{\left(\sqrt{4t^2 + (\omega-i \mu)^2} + \sqrt{4t^2 + (\omega+ i \mu)^2} -2 |\omega| \right)}{4 |\omega|} + \mathcal{O}(\Delta^3) \,, \nonumber \\
G(i \omega) &=& G_0 (i \omega) + \mathcal{O}(\Delta^2) \,.
\eea
Note that $F(i \omega)$ is a real and even function of $\omega$ along the imaginary frequency axis. However, neither $F$ nor $G$ are analytic at $\omega=0$. Similarly, we can see that $G(-i \omega) = G^\ast (i \omega)$, and for $\mu=0$ $G (i \omega)$ is purely imaginary, with $G(-i\omega) = - G(i \omega)$.

\begin{figure}[t]
    \centering
    \subfloat[]{\includegraphics[width=0.45\textwidth]{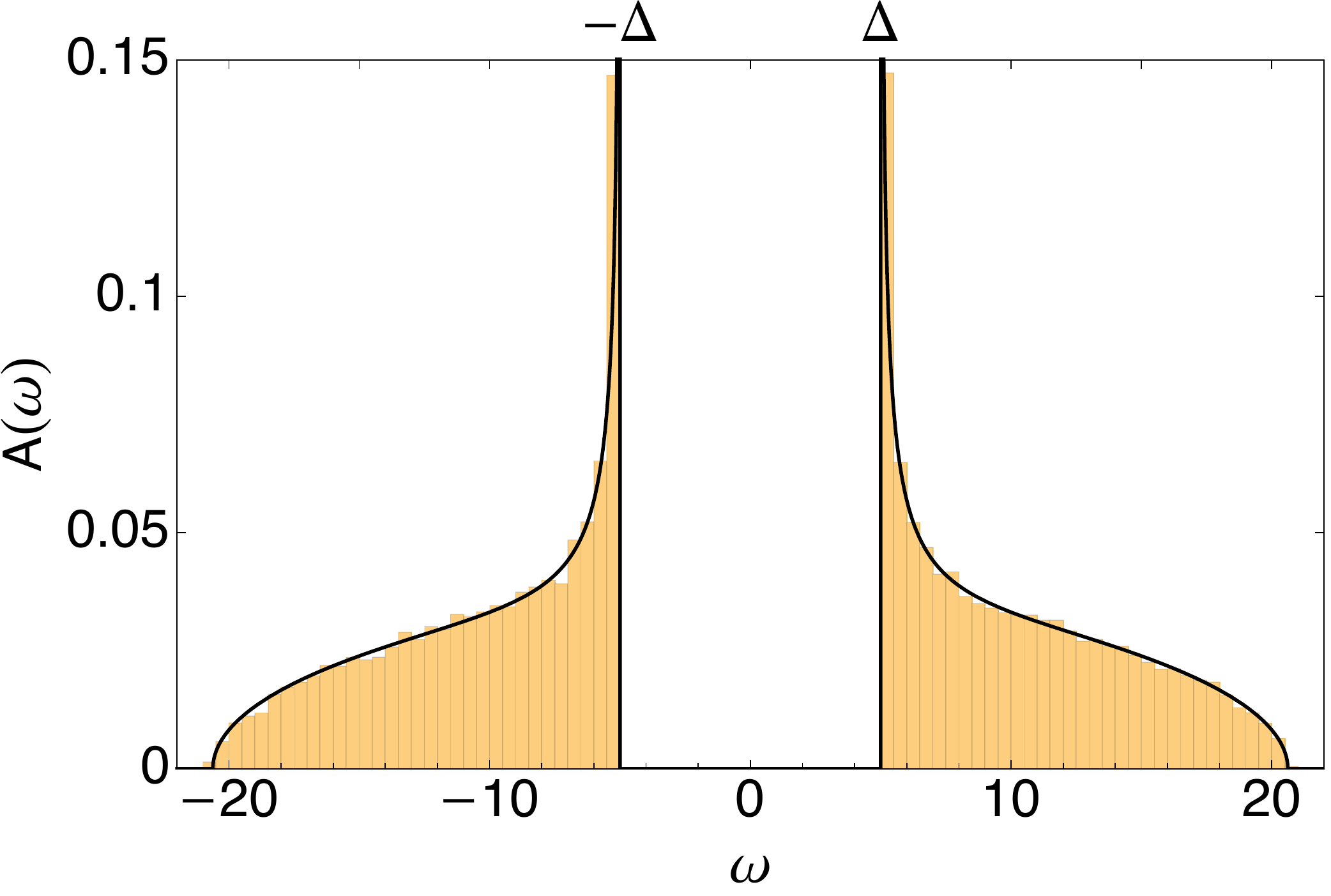}}~~~~
    \subfloat[]{\includegraphics[width=0.45\textwidth]{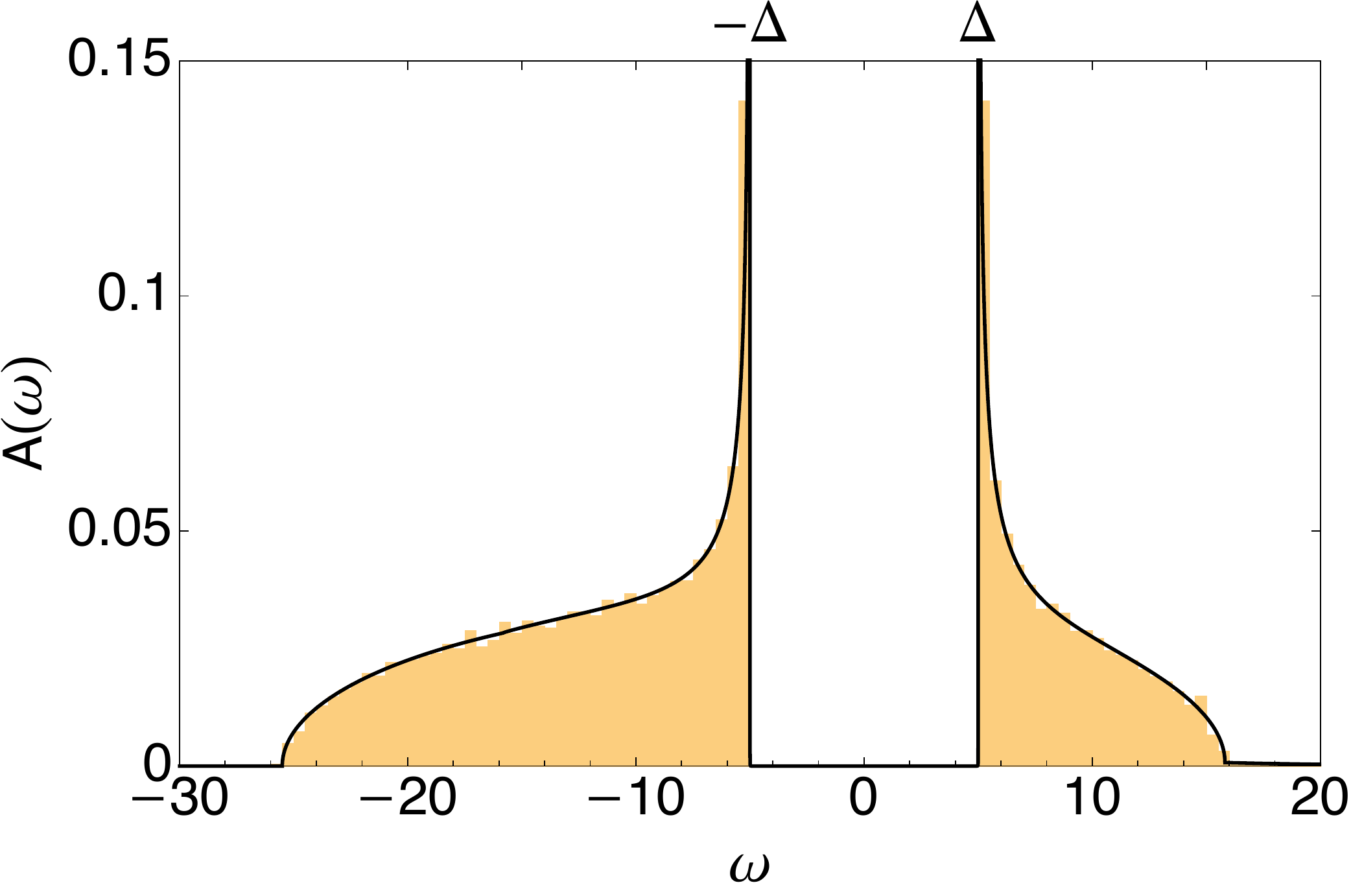}}~~~~\\
    \subfloat[]{\includegraphics[width=0.45\textwidth]{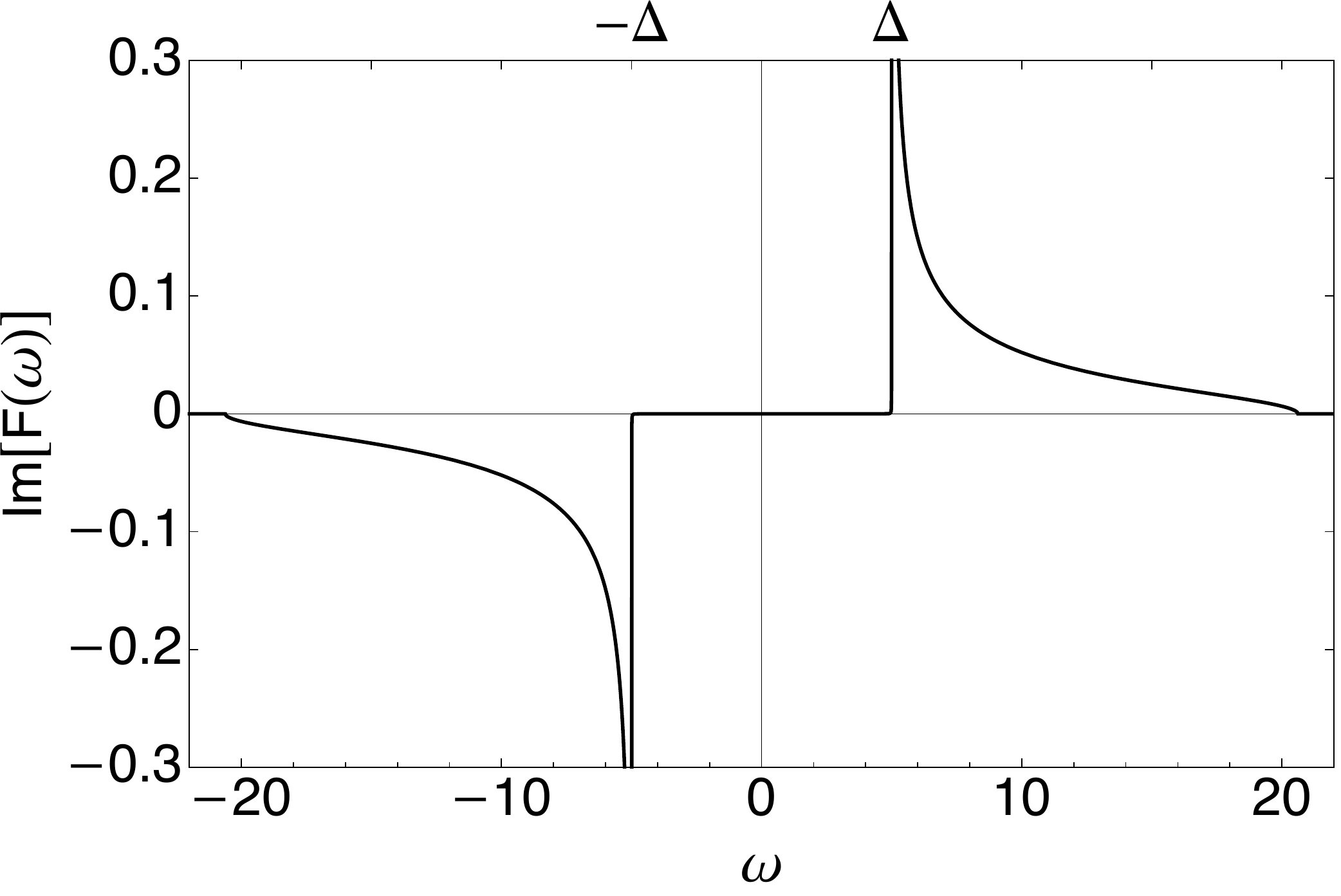}}~~~~
    \subfloat[]{\includegraphics[width=0.45\textwidth]{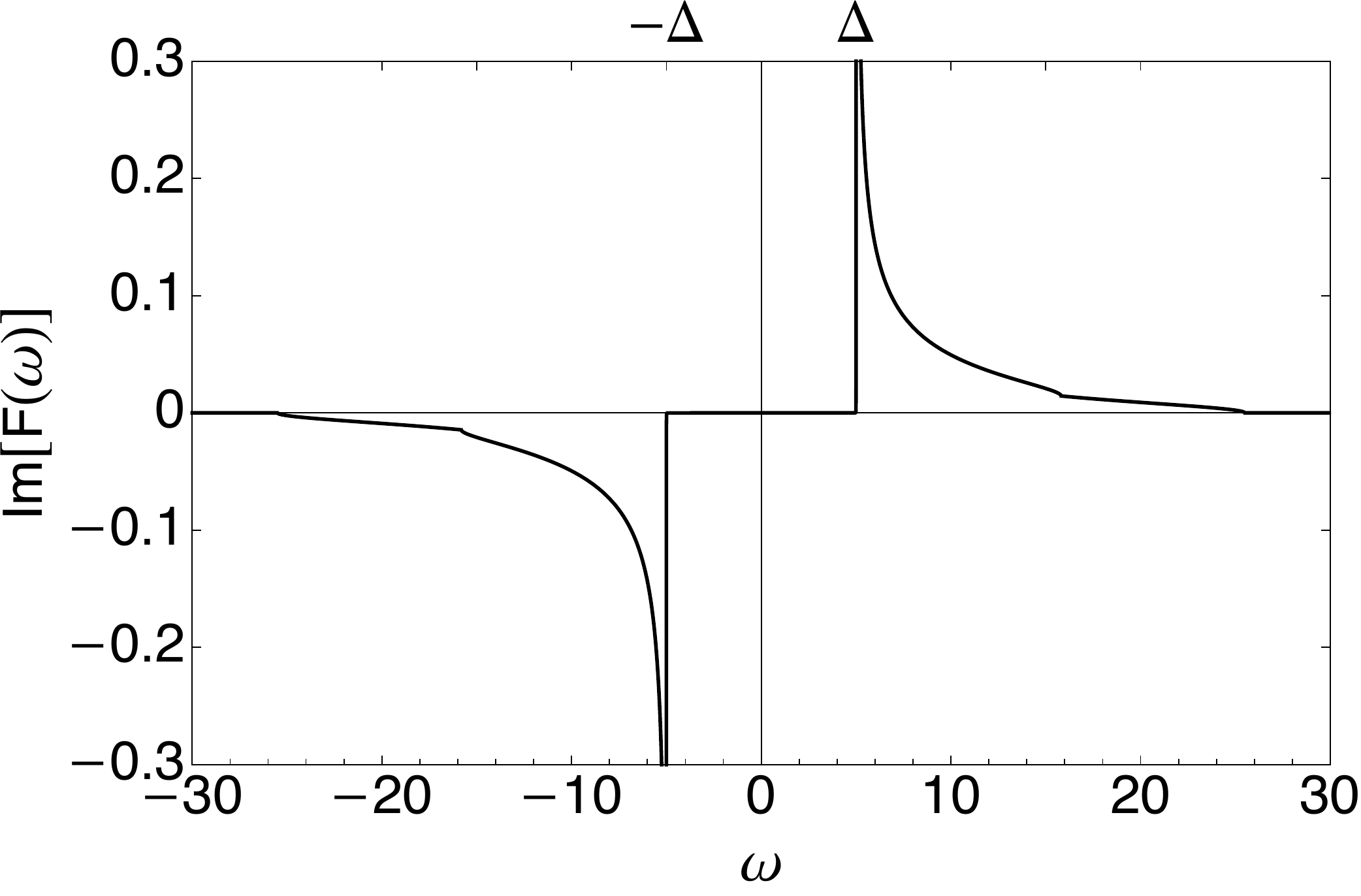}}~~~~
    \caption{ (a) The spectral function, $A(\omega)$, of the normal Green's function in the SC phase at a fixed $\Delta$ for the particle-hole symmetric case ($\mu=0$). The solid line is exact solution to the saddle point equations Eqs. (\ref{eall}), and the yellow bars are obtained by averaging exact diagonalizations of random instances of Eq. (\ref{eq:bcs_ham}). (b) Same as (a) but $\mu=5$. 
    (c) Imaginary part of the anomalous Green’s function $F(\omega)$ in the SC phase at a fixed $\Delta$ and $\mu=0$. (d) Same as (c) with $\mu=5$.  
    In all the plots, $t=10$ and $\Delta=5$. 
    }
    \label{fig:spec_bcs}
\end{figure}

At $\mu=0$, the exact solution of the saddle-point equations in (\ref{eall}) is 
\bea
G(i \omega) &=& - \frac{i \omega}{2t^2} \left( \frac{\sqrt{\omega^2 + 4t^2 + \Delta^2}}{\sqrt{\omega^2 + \Delta^2}} - 1 \right) \,, \nonumber \\
F(i \omega) &=& \frac{\Delta}{2t^2} \left( \frac{\sqrt{\omega^2 + 4t^2 + \Delta^2}}{\sqrt{\omega^2 + \Delta^2}} - 1 \right)\,.
\eea
Analytic continuation gives the spectral function, $A(\omega)\equiv -\frac{1}{\pi} \Im G(\omega+i\delta)$,
\beq
A(\omega)=\frac{|\omega|}{2\pi t^2}\frac{\sqrt{4t^2+\Delta^2-\omega^2}}{\sqrt{\omega^2-\Delta^2}}\,,\quad \Delta<|\omega|<\sqrt{\Delta^2+4t^2}\,.
\label{eq: ac}
\eeq
The spectral function is plotted in Fig. \ref{fig:spec_bcs}a, along with the numerical results obtained by exact diagonalization of random realizations of the Hamiltonian in Eq. (\ref{eq:bcs_ham}).
As expected, the gap is centered at $\omega=0$, between $\Delta$ and $-\Delta$. It is also straightforward to obtain the imaginary part of the retarded anomalous Green's function, which is shown in Fig.~\ref{fig:spec_bcs}c.
For $\mu\neq 0$ an analytic solution is no longer possible, and we show numerical results in Figs.~\ref{fig:spec_bcs}b,d.


\section{Model}
\label{sec:mod}

Having discussed the basic set-up we are now ready to discuss our model. 
To the random Hubbard model considered in the previous section, we will now add random and all-to-all spin exchange and Cooper-pair hopping terms. So the full Hamiltonian is
\begin{align}
\label{eq:Ham}
H &= H_{tU} + H_J + H_L \,, \\
\label{eq:H_tU}
H_{tU} &= - \frac{1}{\sqrt{N}} \sum_{i<j} t_{ij} \left( c_{i \alpha}^\dagger c_{j}^{\alpha} + c_{j \alpha}^\dagger c_{i}^ {\alpha} \right) + \sum_i \left[
- \mu c_{i \alpha}^\dagger c_{i}^{\alpha} + \frac{U}{2M} \left| \mathcal{J}^{\alpha\beta} c_{i \alpha}^\dagger c_{i \beta}^{\dagger} \right|^2  \right] \,, \\
\label{eq:H_J}
H_{J} &= \frac{1}{\sqrt{NM}} \sum_{i<j} J_{ij} \, c_{i \alpha}^\dagger c_i^{\beta} c_{j \beta}^\dagger c_j^{\alpha}  \,, \\
\label{eq:H_L}
H_{L} &= - \frac{1}{2\sqrt{NM}} \sum_{i<j} L_{ij} \mathcal{J}^{\alpha\beta} \mathcal{J}_{\gamma\delta} \left[ c_{i \alpha}^\dagger c_{i \beta}^\dagger c_{j}^{\gamma} c_j^{\delta} + c_{j \alpha}^\dagger c_{j \beta}^\dagger c_{i}^{\gamma} c_i^{\delta}\right] \,.
\end{align}
Recall that we have solved $H_{tU}$ in Sec. \ref{sec:BCS}.
$H_J$ describes the exchange interaction of the
original SY model \cite{SY92}, while $H_{L}$ describes the random Cooper-pair hopping. 
In the above Hamiltonian, $J_{ij}$ are real random numbers with zero mean value and root-mean-square value of $J$. Similarly, $L_{ij}$ can be either real or complex random numbers with zero mean value and root-mean-square value of $L$.

For clarity, let us consider the contribution of individual terms in the Hamiltonian in Eq. (\ref{eq:Ham}). The first term, $H_{tU}$, in Eq. (\ref{eq:H_tU}) was already dealt with in Sec. \ref{sec:BCS}. Next, let us consider the contribution of $H_{J}$ in Eq. (\ref{eq:H_J}) to the action of the full Hamiltonian. 
After averaging over Gaussian random variable $J_{ij}$ the resulting action is
\beq
\overline{\mathcal{S}}_J = - \frac{J^2}{4 NM} \int d \tau d \tau' \left| \sum_{i} c_{i \alpha}^\dagger (\tau) c_i^{\beta} (\tau) c_{i \gamma}^\dagger (\tau') c_i^{\delta} (\tau') \right|^2  \,.
\label{eq:Sb_J} 
\eeq
In the large $M$ limit, we can use an identity analogous to Eq. (\ref{Me1}),
\beq
c_{\alpha}^\dagger (\tau) c^{\beta} (\tau) c_{\gamma}^\dagger (\tau') c^{\delta} (\tau') \Rightarrow \delta_\alpha^\delta \delta^\beta_\gamma G(\tau, \tau') G(\tau', \tau) + \mathcal{J}^{\beta\delta} \mathcal{J}_{\alpha\gamma} F^\ast (\tau, \tau') F(\tau, \tau') \,.
\label{Me2}
\eeq
Here we have dropped factorizations associated with equal-time Green's functions.
Then the contribution to the action from the $H_J$ term is $\mathcal{S}_J$ with,
\beq
\frac{\mathcal{S}_J}{NM} = - \frac{J^2}{4} \int d \tau d \tau' \left( \left[ G(\tau, \tau') G(\tau', \tau) \right]^2 + \left| F(\tau, \tau') F(\tau', \tau) \right|^2 \right) \,. \label{eq:SJ}
\eeq

Finally, let us consider the contribution from the random Cooper-pair hopping term, $H_{L}$, in Eq. (\ref{eq:H_L}). 
Averaging over real Gaussian random variable $L_{ij}$ yields the action,
\bea
\overline{\mathcal{S}}_L &=& - \frac{L^2}{8 NM} \int d \tau d \tau' \mathcal{J}^{\alpha\beta} \mathcal{J}^{\mu\nu} \mathcal{J}_{\gamma\delta} \mathcal{J}_{\rho\sigma} \Biggl[ \nonumber \\
&~&~~~~~~~~~~~ \left( \sum_i c_{i \alpha}^\dagger (\tau)  c_{i \beta}^\dagger (\tau) c_i^\rho (\tau') c_i^\sigma (\tau') \right)\left( \sum_j c_{j \mu}^\dagger (\tau')  c_{j \nu}^\dagger (\tau') c_j^\gamma (\tau) c_j^\delta (\tau) \right) \nonumber \\
&~&~~~~~~~~+  \left( \sum_i c_{i \alpha}^\dagger (\tau)  c_{i \beta}^\dagger (\tau) c_{i \mu}^\dagger (\tau') c_{i \nu}^\dagger (\tau') \right)\left( \sum_j c_{j}^{\gamma} (\tau)  c_{j}^{\delta}  (\tau) c_j^\rho (\tau') c_j^\sigma (\tau') \right)\Biggr] \,.
\label{eq:Sb_L}
\eea
Note that the last term would be absent for complex $L_{ij}$.
Now, we use large $M$ identities similar to Eqs. (\ref{Me1}) and (\ref{Me2}), again dropping equal-time factorizations,
\bea
c_{\alpha}^\dagger (\tau)  c_{\beta}^\dagger (\tau) c^\rho (\tau') c^\sigma (\tau') & \Rightarrow & \left(\delta_{\alpha}^\sigma \delta_\beta^{\rho} -\delta_{\alpha}^\rho \delta_\beta^{\sigma} \right) \left[G(\tau, \tau') \right]^2 \,, \nonumber \\
c_{\alpha}^\dagger (\tau)  c_{\beta}^\dagger (\tau) c_{\mu}^\dagger (\tau') c_{\nu}^\dagger (\tau') & \Rightarrow & 
\left( \mathcal{J}_{\alpha\nu} \mathcal{J}_{\beta \mu} - \mathcal{J}_{\alpha\mu} \mathcal{J}_{\beta \nu} \right)\left[F^\ast (\tau, \tau') \right]^2 \,.
\label{Me3}
\eea
The contribution of the $H_{L}$ term to the action is $\mathcal{S}_{L}$ with,
\beq
\frac{\mathcal{S}_L}{NM} = - \frac{L^2}{4} \int d \tau d \tau' \left( \left[ G(\tau, \tau') G(\tau', \tau) \right]^2 + \left| F(\tau, \tau') F(\tau', \tau) \right|^2 \right) \,,
\label{eq:SL}
\eeq
having the same form as $\mathcal{S}_J$ in Eq. (\ref{eq:SJ}).

So finally, the action corresponding to the full Hamiltonian in Eq. (\ref{eq:Ham}) is 
\beq
\mathcal{S} = \mathcal{S}_0 + \mathcal{S}_1 + \mathcal{S}_J + \mathcal{S}_L \,,
\label{eq:S_full}
\eeq
with the terms $\mathcal{S}_0$ and $\mathcal{S}_1$ quoted in Eqs. (\ref{S0}) and (\ref{S1}) respectively, while the terms $\mathcal{S}_J$ and $\mathcal{S}_L$ are shown in Eqs. (\ref{eq:SJ}) and (\ref{eq:SL}) respectively.

Putting everything together, the final saddle-point equations for the normal and anomalous equations are
\begin{align} 
\label{eq:Gsig}
G_\Sigma (i\omega) &\equiv  \frac{1}{i \omega +\mu - \Sigma (i\omega)} \,, \\
\label{eq:sig}
\Sigma (\tau, \tau') &= t^2 G (\tau, \tau') - (J^2 + L^2) G^2 (\tau, \tau') G(\tau', \tau) \,, \\
\label{eq:Giw}
G(i \omega) &= \frac{\left[G_\Sigma (-i\omega) \right]^{-1}}{|\Phi(i\omega)|^2 + \left[G_\Sigma (i\omega) G_\Sigma (-i\omega)\right]^{-1}} \,, \\
\label{eq:delta}
\Delta &= -U T \sum_\omega \frac{\Phi(i\omega)}{|\Phi(i\omega)|^2 + \left[G_\Sigma (i\omega) G_\Sigma (-i\omega)\right]^{-1}} \,, \\
\label{eq:F}
F(i\omega) &= \frac{\Phi(i\omega)}{|\Phi(i\omega)|^2 + \left[G_\Sigma (i\omega) G_\Sigma (-i\omega)\right]^{-1}} \,, \\
\label{eq:phi}
\Phi (\tau, \tau') &= - U F(\tau, \tau) \delta(\tau - \tau') + t^2 F(\tau, \tau') + (J^2 + L^2) F^2 (\tau, \tau') F^\ast (\tau', \tau) \,. 
\end{align}
Note that Eqs. (\ref{eq:sig}) and (\ref{eq:phi}) generalize the expressions in Eq. (\ref{saddle1}) upon the inclusion of the spin exchange and Cooper-pair hopping terms. 


\section{Numerical solutions} 
\label{sec:num}

We shall now solve the the saddle-point equations (Eqs. (\ref{eq:Gsig}-\ref{eq:phi})) at finite temperature and obtain the normal-state as well as SC solutions. For simplicity and clarity, we will focus on the $\mu=0$ half-filling case, but the results are qualitatively similar for non-zero $\mu$ as we show at the end of this section. Hence, unless otherwise stated $\mu=0$ throughout this section. We introduce the notation  $\widetilde{J}=\sqrt{J^2+L^2}$ since the interactions $J$ and $L$ are on equal footing in the large-$M$ limit, as seen from Eqs. (\ref{eq:sig}) and (\ref{eq:phi}). Furthermore, we will parameterize the hopping $t$ and interaction $\widetilde{J}$ as 
\begin{equation}
\label{eq:theta}
t = R \cos\theta \,, ~~~
\widetilde{J} = R \sin\theta \,,
\end{equation}
where $R = \sqrt{t^2 + \widetilde{J}^2}$, and the parameter $\theta \in [0,\pi/2]$ tunes between FL ($\theta=0$) and SYK-NFL ($\theta=\pi/2$) limits. We will discuss results for different relative strengths with respect to $U$, i.e., different ratios $R/|U|$.

We solve the saddle-point equations, Eqs. (\ref{eq:Gsig})-(\ref{eq:phi}), on the imaginary (Matsubara) frequency axis at finite temperature. The strategy is as follows. We first start with a  
free fermion normal Green's function, $G(i\omega_n)=(i\omega_n+\mu)^{-1}$, and a randomly chosen real function $F(i\omega_n)$, and iterate until we find a converged solution for the normal and anomalous Green's functions. The SC order parameter, $\Delta(T)=-U\mathcal{J}_{\alpha\beta} \langle c^{\alpha}c^\beta \rangle$, is then determined as a function of temperature. It is finite at low temperatures in the superconducting phase, and it vanishes in the normal state at higher temperature. The superconducting critical temperature, $T_{sc}$, is thus determined numerically using $\Delta(T\to T_{sc}^{-}) \rightarrow 0$. We will use the notation $\Delta_0 \equiv \Delta(T\to 0)$.

In both the normal and SC phases we also compute the spectral function. The spectral function is obtained by numerical analytic continuation of Matsubara Green's functions to the real frequency axis. More details regarding numerical analytic continuation are discussed in Appendix \ref{sec:nac}.

\subsection{Normal State}
\label{sec:ns}

The normal-state equations with $\Delta = 0$ and $F=0$ are the same as those in Refs.~\cite{PG98} and~\cite{Balents17}. As stated earlier, in our model we tune the parameter $\theta$, defined in Eq. (\ref{eq:theta}), to go from FL to NFL normal states. At any given temperature $T$, the normal state is FL like for $\theta \lesssim \theta_{coh}$ and NFL-like for $\theta \gtrsim \theta_{coh}$, where $\theta_{coh}$ is defined by $T \sim T_{coh} = t^{2}/\widetilde{J} = R \cos\theta_{coh} \cot\theta_{coh}$. 

\begin{figure}
    \centering
    \subfloat[]{\includegraphics[width=0.45\textwidth]{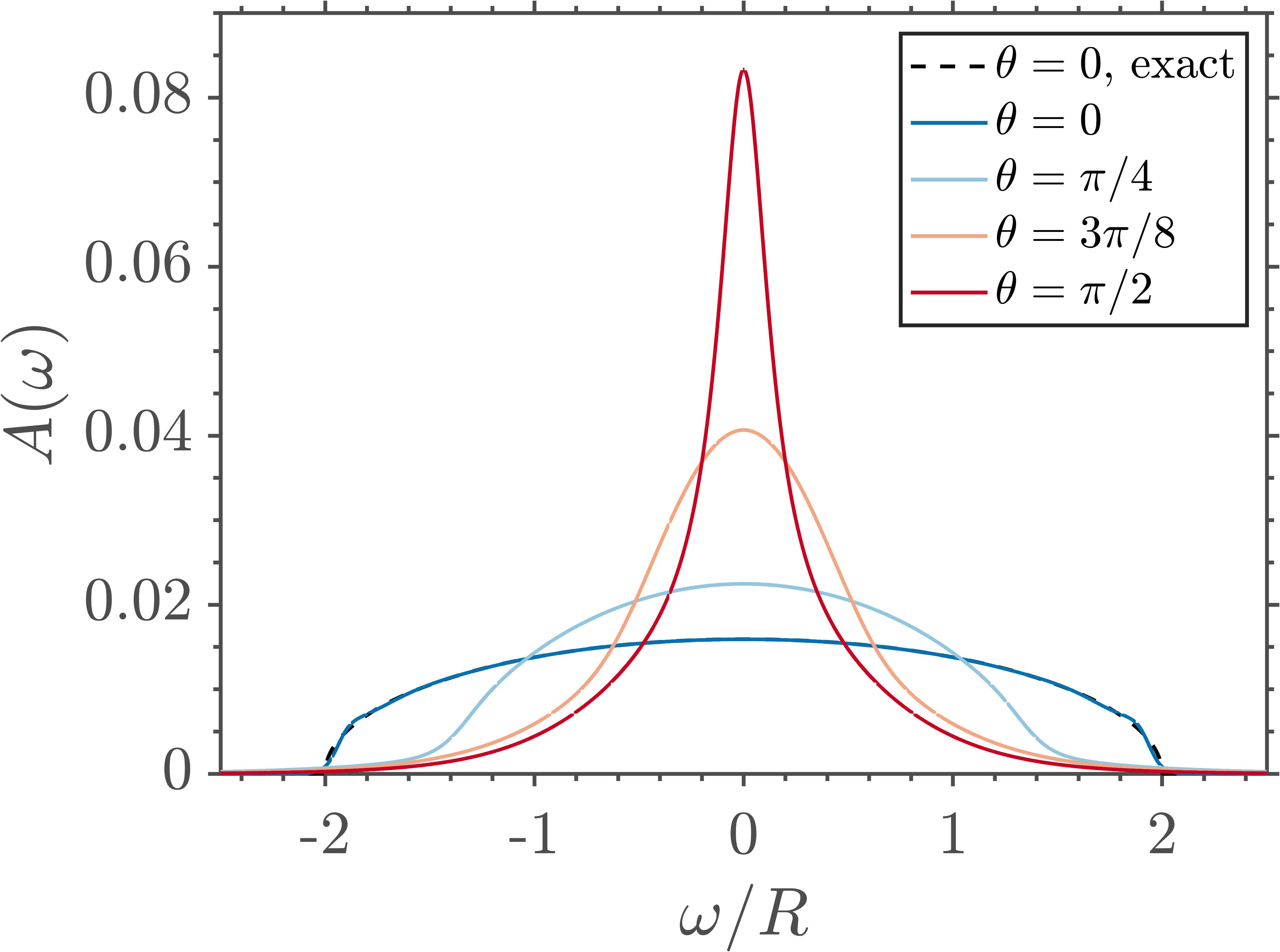} }~~~~ 
    \subfloat[]{\includegraphics[width=0.5\textwidth]{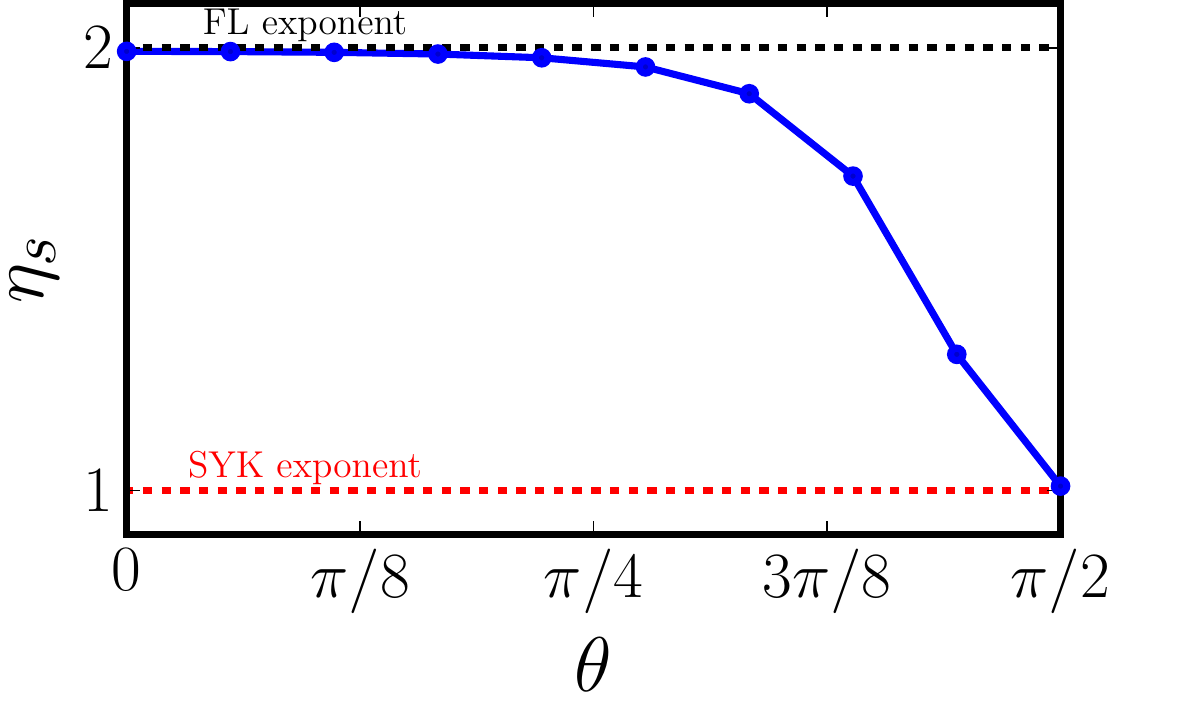}}
    \caption{(a) The normal-state spectral function $A(\omega)$ for different values of $\theta$ at $\mu=0$, and $R/|U|=2$.  The dashed line is the exact semi-circle solution for $\theta=0$, as obtained in Eq. (\ref{Gnormal}). 
    (b) Effective spin exponent as a function of $\theta$ in the normal state at $R/|U|=2$. The spin exponent takes the expected FL value for low $\theta$, while it approaches the SYK value for larger $\theta$.} 
    \label{fig:Aw_ns}
\end{figure}

In Fig. \ref{fig:Aw_ns} (a) we show the spectral function in the normal state. For the FL-like phase (smaller $\theta$) we see the expected semi-circular spectral function, whereas for a NFL-like phase (larger $\theta$) a pronounced peak at $\omega=0$ is seen. This is consistent with earlier results obtained for a similar random model in Ref. \cite{PG98}. 

Also, note that the FL-like normal state ($\theta < \theta_{coh}$) has the usual $T^2$ dependence of resistivity, while the NFL state ($\theta > \theta_{coh}$) has a linear-in-$T$ resistivity. This is similar to the results obtained in Refs. \cite{Balents17, PG98}.

The cross-over between the FL and NFL normal states can be further characterized by looking at the effective spin exponent ($\eta_{s}$), which is shown in Fig. \ref{fig:Aw_ns} (b). This exponent is extracted from the dynamical susceptibility, $\chi''(\omega)$, which is the imaginary part of the spin correlation. In Appendix \ref{sec:etaS} we discuss the details related to the evaluation of the spin exponent $\eta_{s}$. Clearly, for lower values of $\theta$ the spin exponent takes the value $\eta_{s} =2$ expected for a disordered FL, while in the limit $\theta \to \pi/2$ it takes the value $\eta_{s} =1$ corresponding to the marginal NFL. In the intermediate $\theta$ region $\eta_{s}$ smoothly interpolates between these extreme values. As expected, this crossover is roughly around $\theta_{coh}$.


\subsection{Superconducting state}
\label{sec:sc}

Before we discuss the numerical results, we first show analytically that SC phase exists at zero temperature for any infinitesimal attractive on-site interaction. The analysis is similar to that presented in Sec. \ref{sec:BCS}. 
We determine the instability to the superconducting state by expanding the action to second order in $F(\iw)$. This leads to the same condition for the instability as Eq. (\ref{eqTc}). But the important difference is that the Green's function now also contains contribution from the exchange interaction terms and satisfies the equations,
\bea
G_0 (i\omega) & = & \frac{1}{i \omega +\mu - \Sigma_{el} (i \omega) - \Sigma_{in} (i\omega)} \,, \nonumber \\
\Sigma_{el} (i \omega) &=&  t^2 G_0 (i \omega) \,, \nonumber \\
\Sigma_{in} (\tau) &=& - (J^2 + L^2) [G_0 ( \tau)]^2 G_0 (-\tau) \,.
\label{sc1}
\eea
Note that we have separated the self energy into an `elastic' part $\Sigma_{el}$, and an `inelastic' part $\Sigma_{in}$. This is useful because $\mbox{Im} \, \Sigma_{in} (\omega \rightarrow 0) = 0$ at $T=0$, and that is not true for the elastic part. 

From Eq.~(\ref{sc1}), we can write a quadratic equation for $G_0 (0)$:
\beq
t^2 [G_0 (0)]^2 - (\mu-\Sigma_{in} (0)) G_0 + 1 =0 \,.
\eeq
An important point is that $\Sigma_{in} (0)$ is real, and so it can be absorbed into $\mu$. This quadratic equation has two roots, and they correspond to $G_0 (i0^+)$ and $G_0 (i0^-)$.  From the formula for the product of the roots of a quadratic equation we can therefore conclude that at $T=0$,
\beq
\lim_{ \omega \rightarrow 0}  G_0 (i \omega) G_0(-i \omega) = \frac{1}{t^2} \,. 
\eeq
So this equation holds even when $J$ or $L$ are non-zero, and the denominator in Eq. (\ref{eqTc}) vanishes. Thus indicating the presence of SC at $T=0$.

\begin{figure}
    \centering
    \subfloat[]{\includegraphics[width=0.45\textwidth]{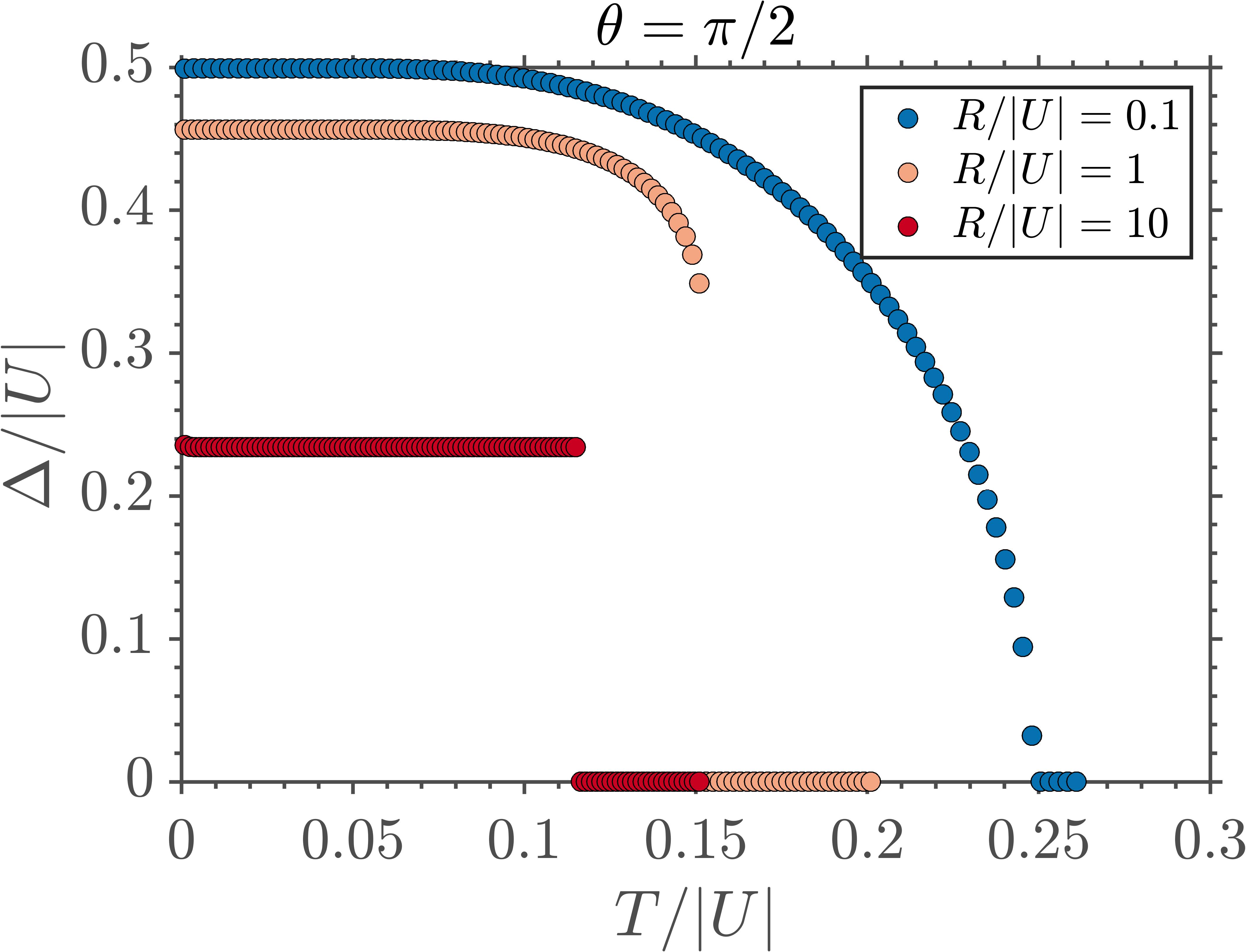}}~~~~
    \subfloat[]{\includegraphics[width=0.45\textwidth]{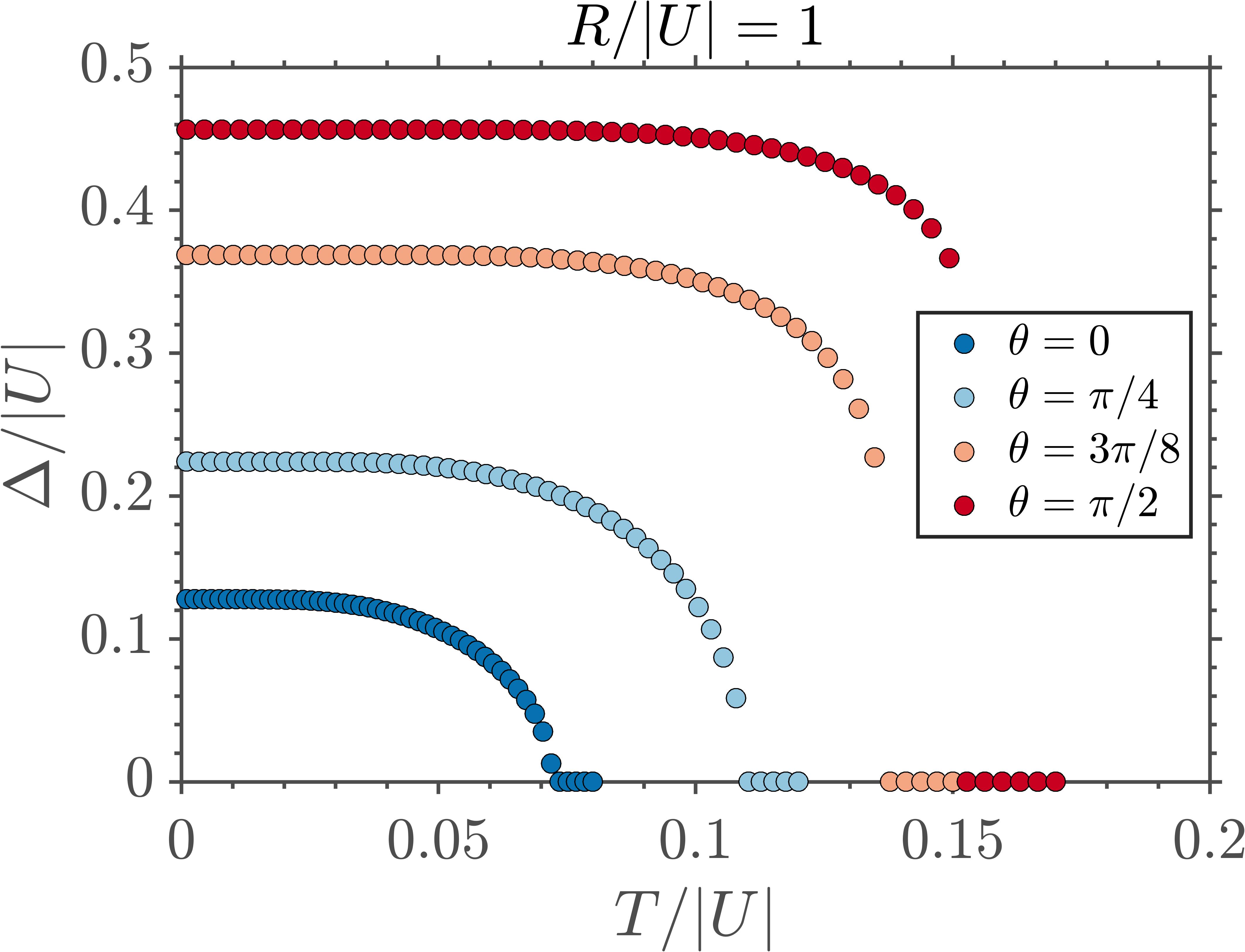}}~~~~
     \caption{ SC order parameter, $\Delta$, as a function of temperature, $T$. 
     (a) SYK-NFL ($\theta=\pi/2$) case with varying $R/|U|$. Note that for larger values of $R/|U|$ the phase transition becomes first order instead of a continuous transition.
     (b) Here $R=|U|$ is fixed and $\theta$ is varied. 
     }
    \label{fig:deltaT}
\end{figure}

Let us now discuss the numerical results obtained by solving the saddle-point equations. For low enough temperature we find a SC solution with a non-zero $\Delta$ and $F(\iw)$. In Fig. \ref{fig:deltaT} we have shown the variation of SC order parameter, $\Delta$, with temperature. It turns out that for small values of $\theta$, i.e., FL-like normal state, the SC-normal state transition is continuous. However, at larger values of $\theta$, the phase transition (SC to NFL) becomes first order for larger values of $R/|U|$ (as seen in Fig. \ref{fig:deltaT} (a)).  Note that although the absolute value of $\Delta$ and $T_{sc}$ depends on the value of $U$, the variation of $\Delta/|U|$ as a function of $T/|U|$ depends only on the ratio $R/|U|$.

\begin{figure}
    \centering
    \subfloat[]{\includegraphics[width=0.45\textwidth]{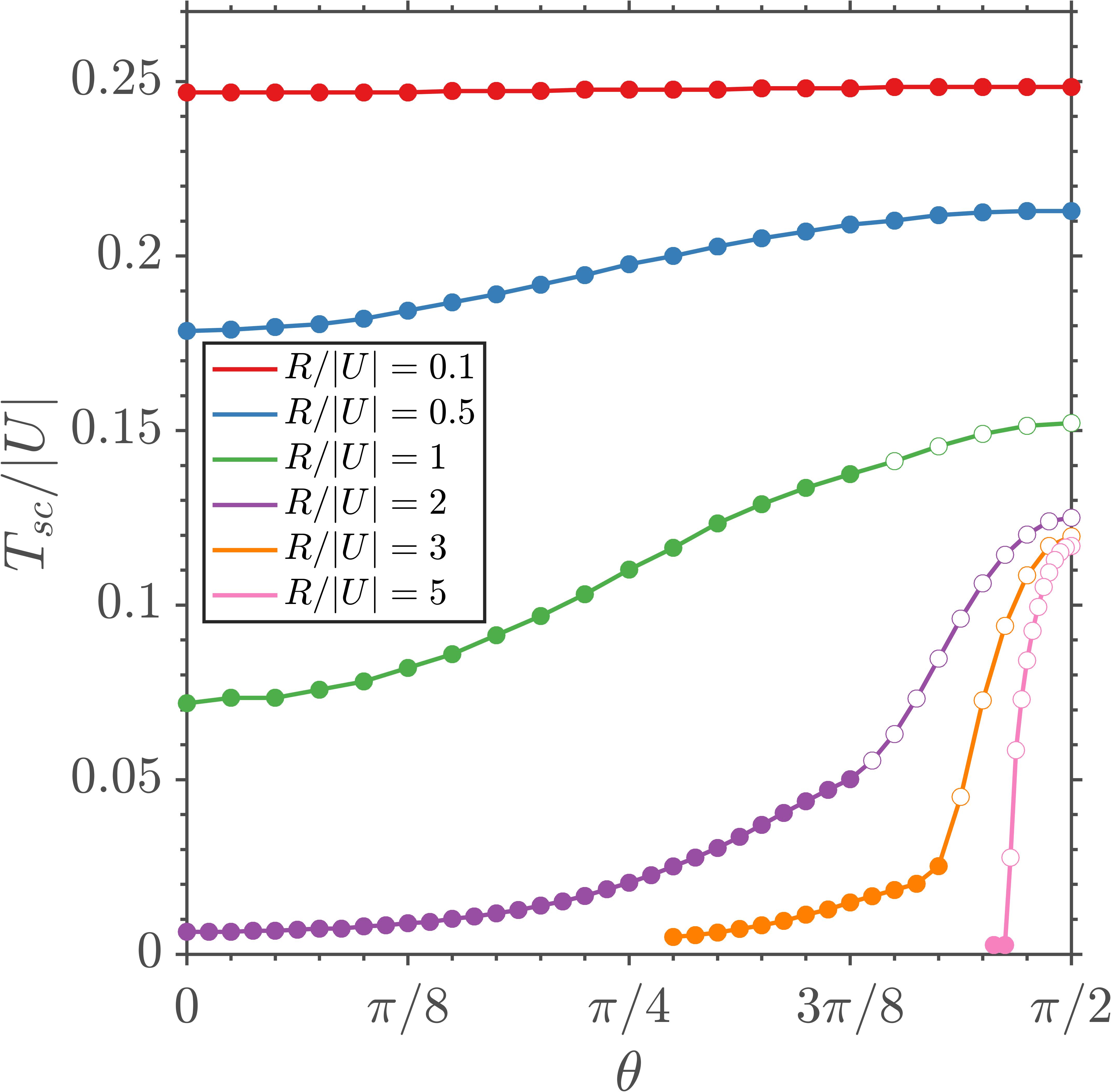}}~~
    \subfloat[]{\includegraphics[width=0.25\textwidth]{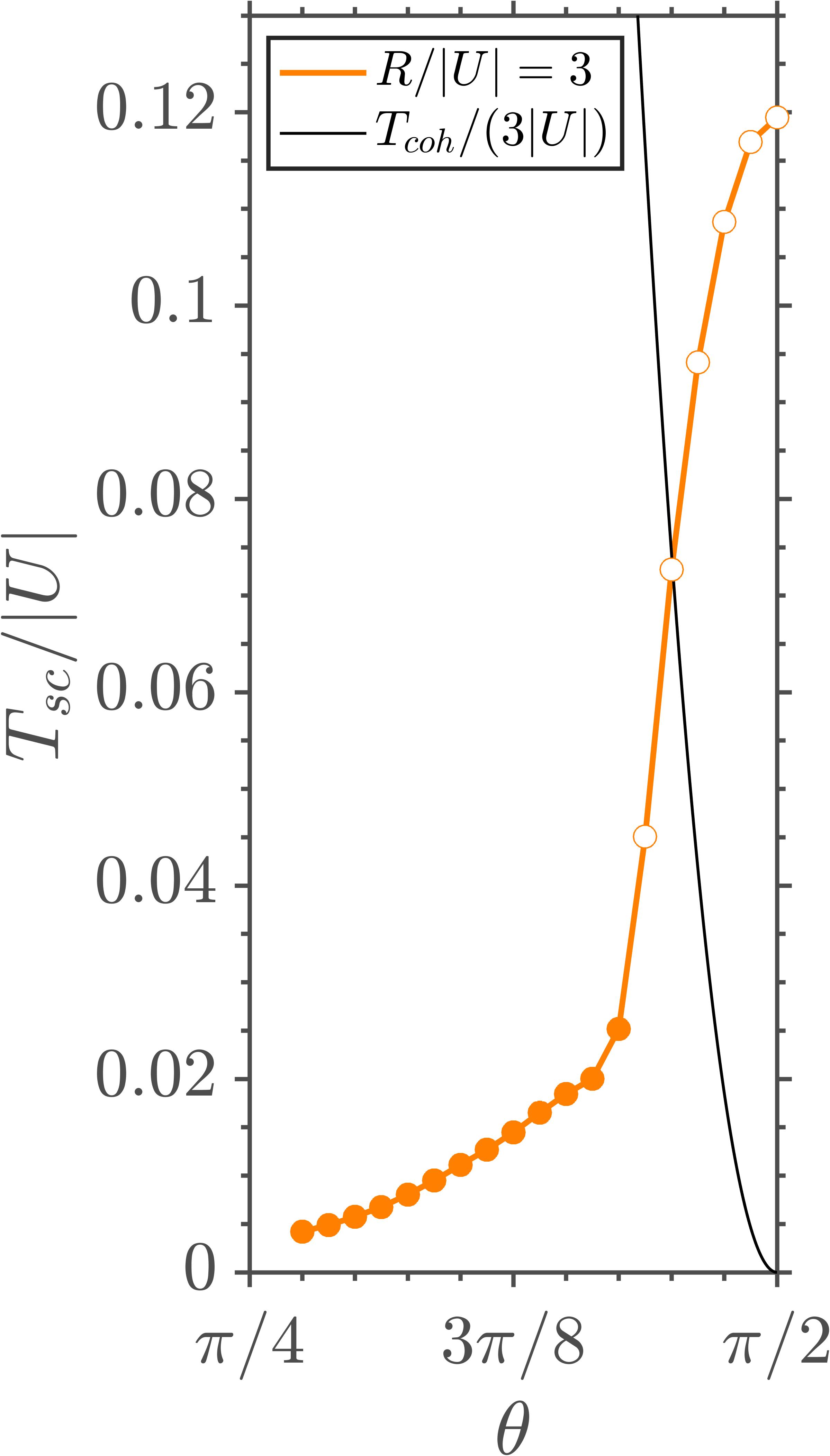}}~~
    \subfloat[]{\includegraphics[width=0.25\textwidth]{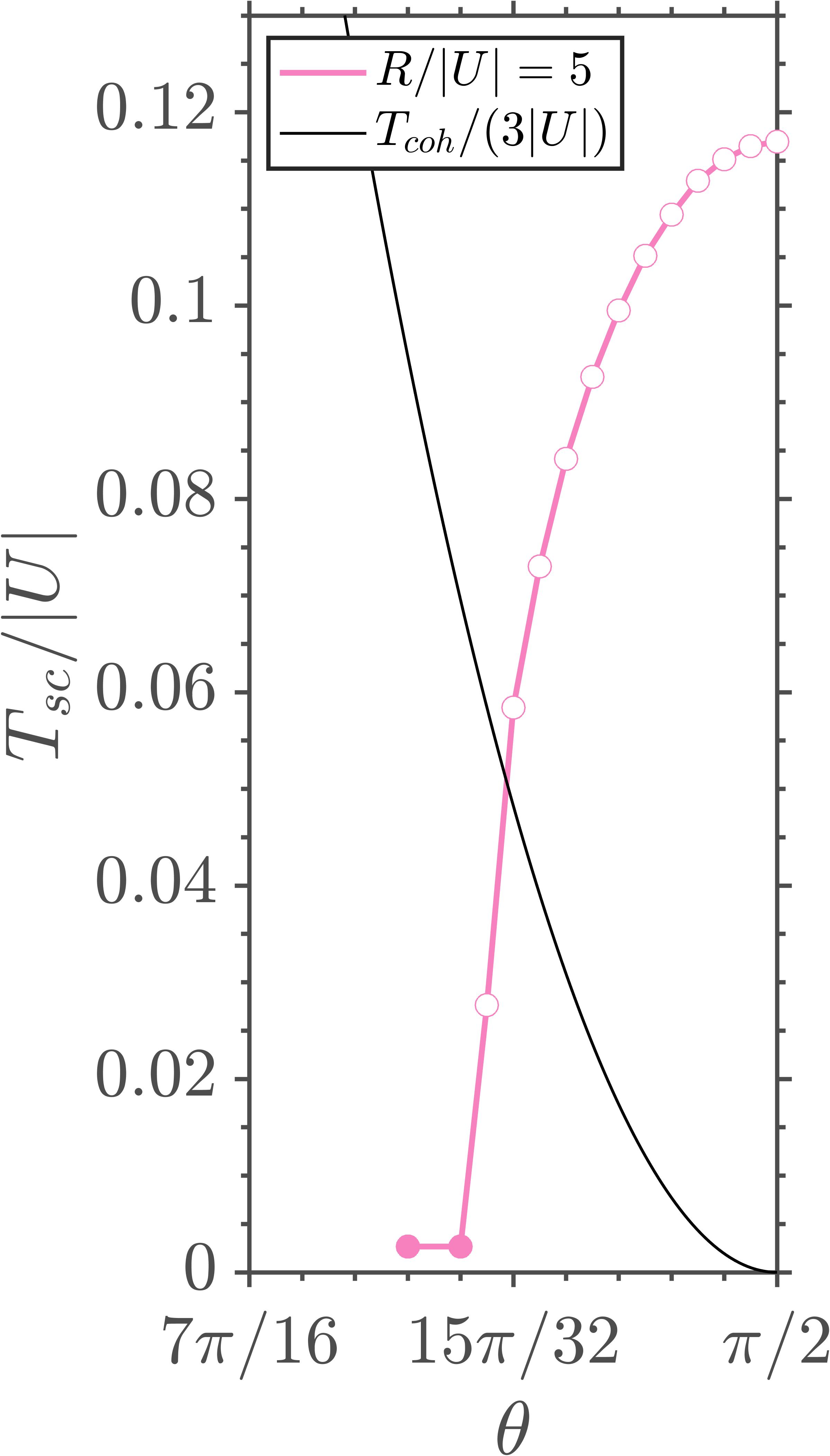}}
    \caption{(a)  The SC transition temperature, $T_{sc}$, as a function of $\theta$ for different values of $R/|U|$.  Qualitatively, the phase transition becomes first order (indicated by open circles) at larger values of $R/|U|$ and $\theta$ instead of a continuous transition (indicated by filled circles). 
    (b) Comparison of $T_{sc}$ and $T_{coh}/3= t^{2}/3J = (1/3)R\cos{\theta}\cot{\theta}$ at $R/|U|=3$. For larger values of $R/|U|$ the transition becomes first order for $\theta \gtrsim \theta_{coh}$. (c) Same as (b) but $R/|U|=5$. }
    \label{fig:Tsc}
\end{figure}
\begin{figure}
    \centering
    \subfloat[]{\includegraphics[width=0.45\textwidth]{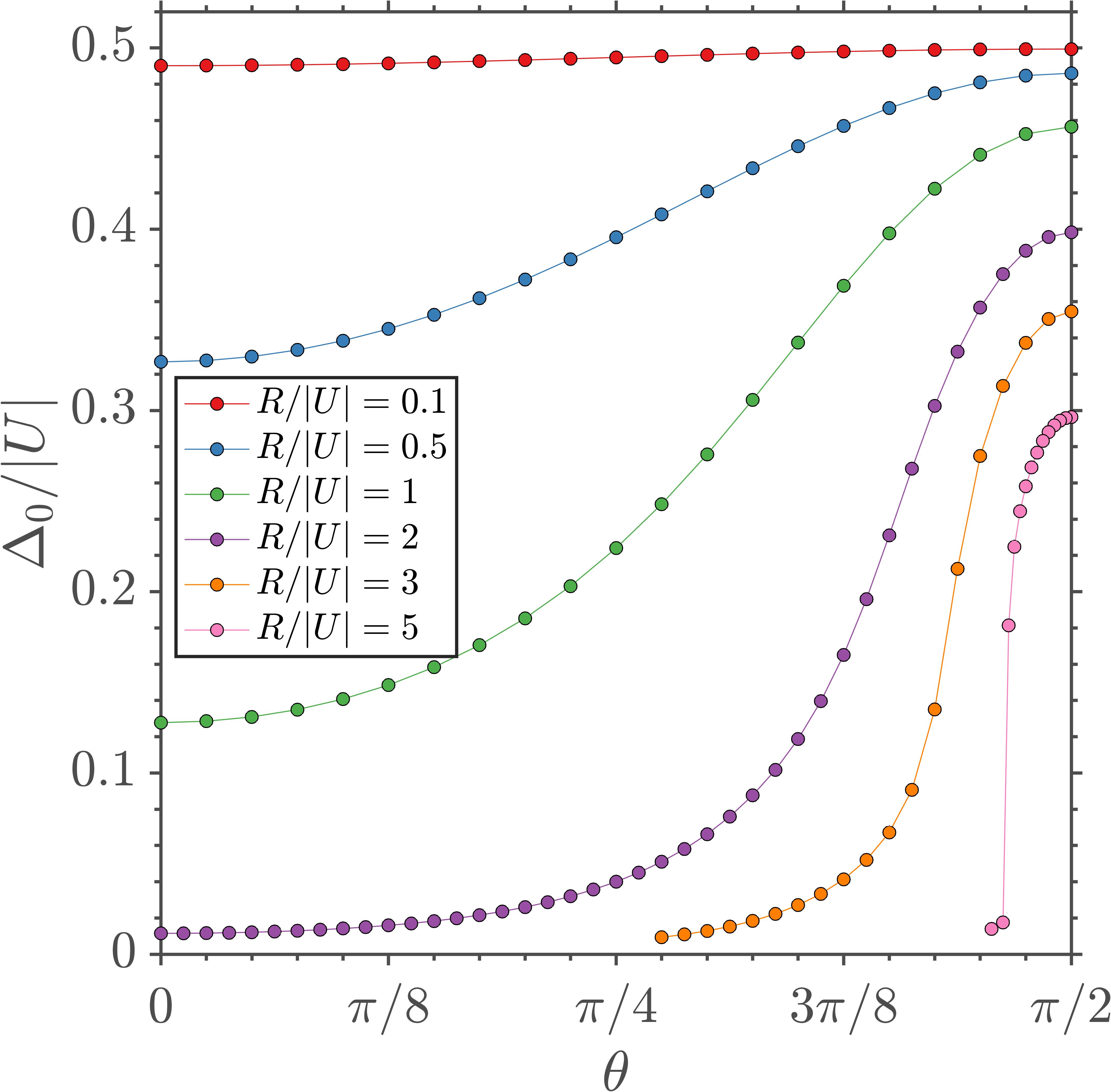}}~~
    \subfloat[]{\includegraphics[width=0.45\textwidth]{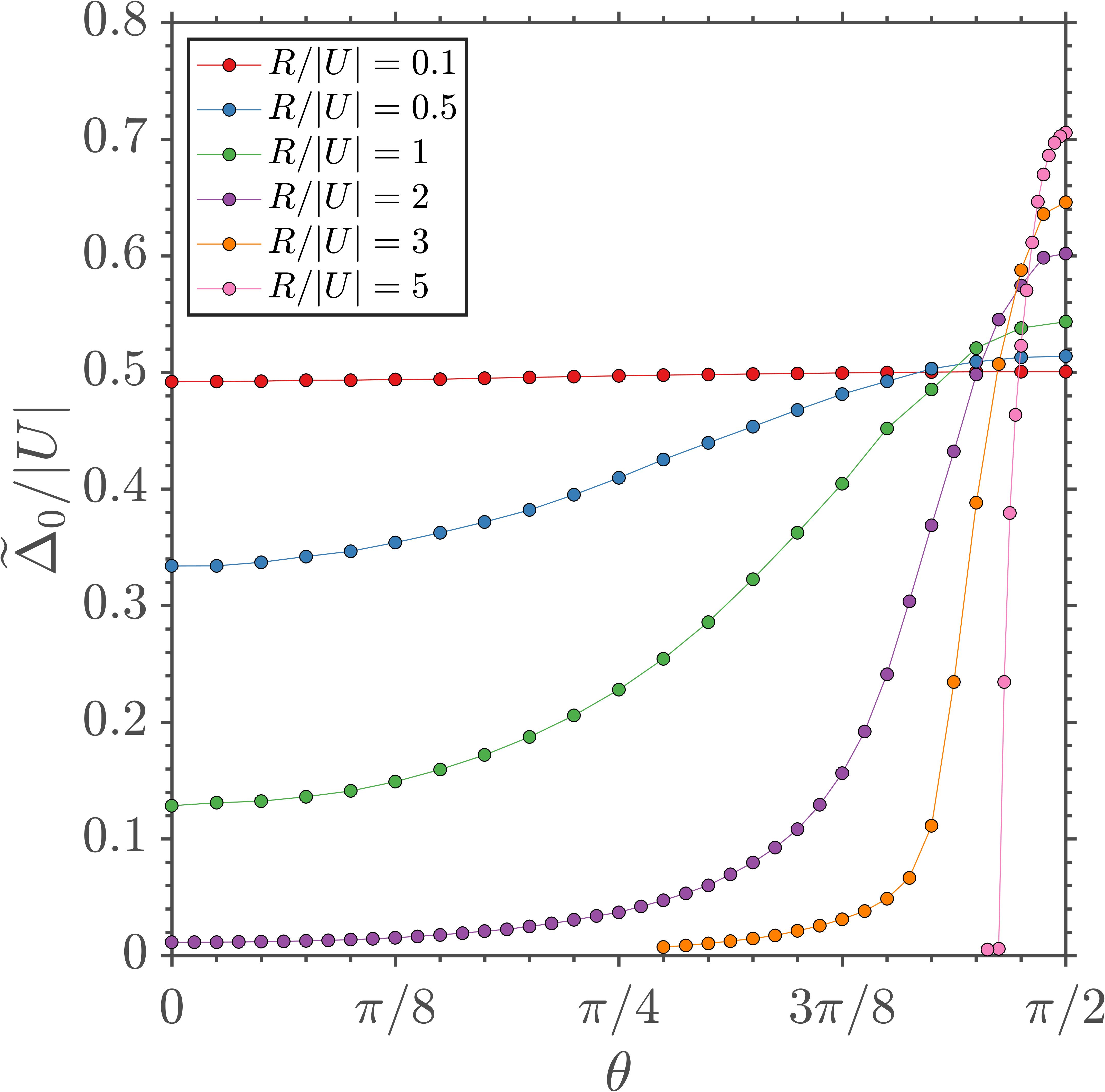}}\\
    \subfloat[]{\includegraphics[width=0.45\textwidth]{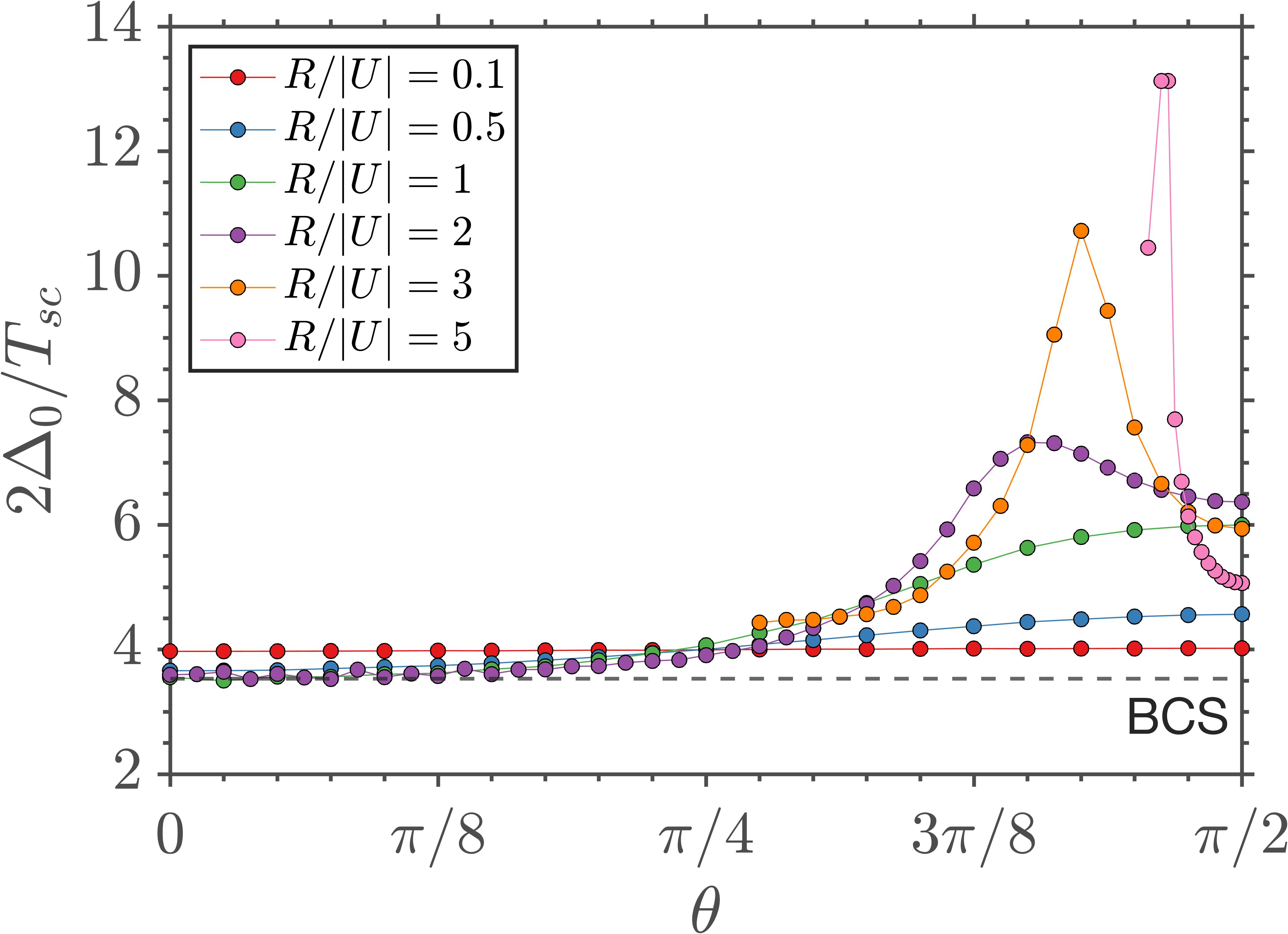}}~~
    \subfloat[]{\includegraphics[width=0.45\textwidth]{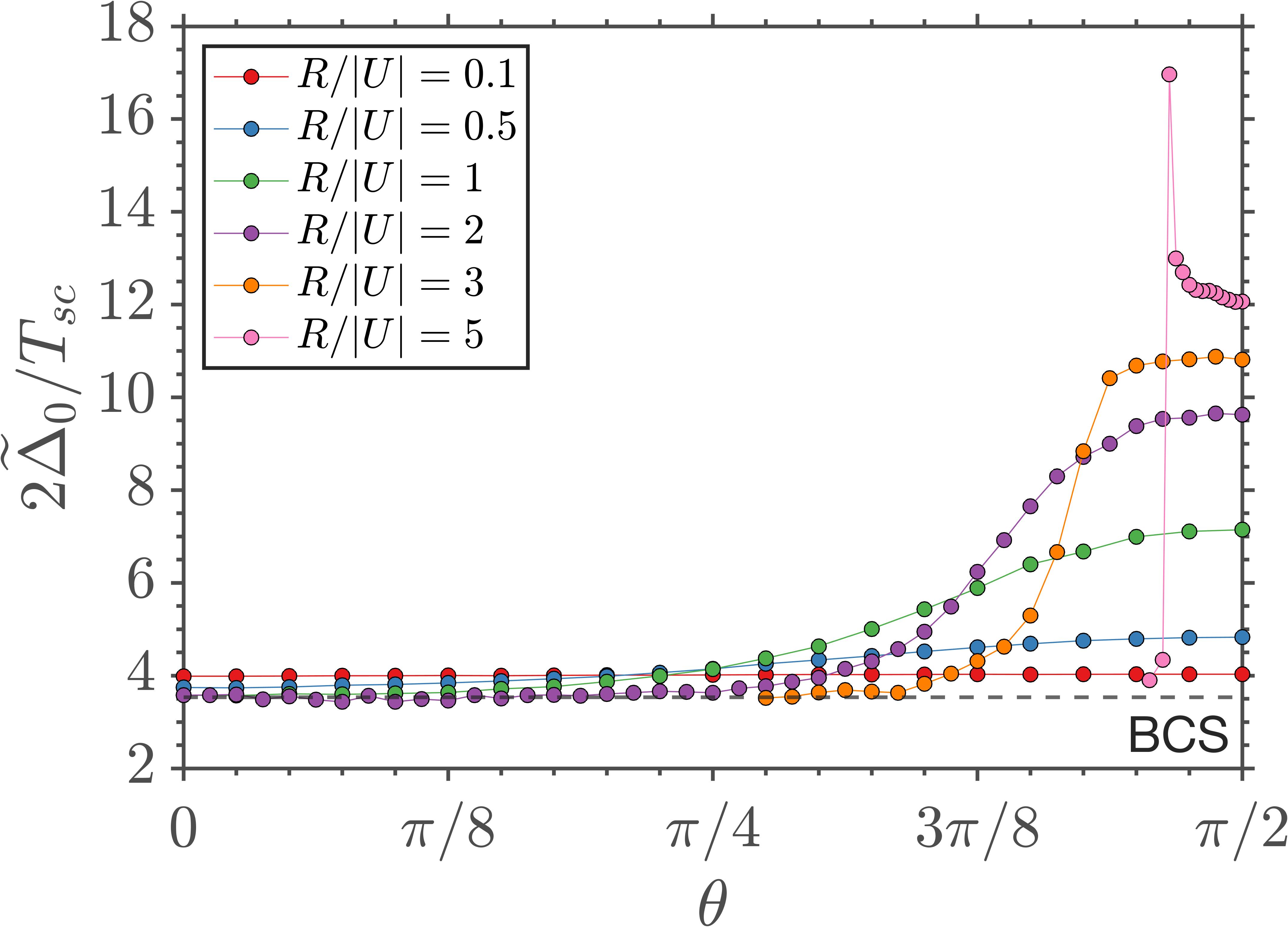}}
    \caption{  (a) The variation of SC order parameter in the zero temperature limit, $\Delta_0$, with $\theta$ for different values of the ratio $R/|U|$. (b) The SC gap observed in the spectral function in the zero temperature limit, $\widetilde{\Delta}_0$, as a function of $\theta$. (c) The ratio $2\Delta_0/T_{sc}$. (d) The ratio $2\widetilde{\Delta}_0/T_{sc}$.  For larger values of $\theta$, these ratios deviate strongly away from the BCS value of $3.53$. As $\theta\rightarrow 0$ and $R/|U| \gg 1$, both $2\Delta_0/T_{sc}$ and  $2\widetilde{\Delta}_0/T_{sc}$ tend to the BCS result. 
    }
    \label{fig:gap}
\end{figure}
In Fig. \ref{fig:Tsc}, we show the variation of SC transition temperature ($T_{sc}$)
as a function of $\theta$ for different values of $R/|U|$. For very large on-site interaction, i.e., for very small $R/|U|$ there is no difference between SC emerging from FL or NFL. This is because in this case both hopping as well as exchange interaction are sub-dominant. However, at larger values of the ratio $R/|U|$, i.e., weaker on-site interaction the SC transition temperature, $T_{sc}$, strongly depends on the nature of the normal state or $\theta$. It is larger for NFL-SC transition (larger $\theta$) as compared to the FL-SC transition (smaller $\theta$). 
The same trend applies to the SC order parameter in the limit of zero temperature, $\Delta_0$, and the SC gap (as obtained from the spectral function) in the $T \to 0$ limit, $\widetilde{\Delta}_0$, as seen in Fig. \ref{fig:gap}. 
Recall that in our model SC phase corresponds to the condensation of doublon, {\it i.e.\/}, the Cooper pairs are on the same site. A single-particle hopping tends to break these pairs and destroy SC. The exchange interaction and Cooper-pair hopping have a very weak effect in destruction of SC. Therefore, $T_{sc}$, $\Delta_0$, and $\widetilde{\Delta}_0$ have very weak dependence on $\theta$ for larger on-site interaction (smaller $R/|U|$), 
as in this case the relative strength of hopping and spin-exchange is unimportant. On the other hand, for weaker on-site interaction the relative strength of hopping, $t$, compared to $\widetilde{J}$ is important. Hence for larger $\theta$ (weaker $t$) SC is more stable leading to a higher $T_{sc}$. 
This is also the reason why the SC-NFL transition becomes first order in nature for larger $R/|U|$.

We have also calculated the ratio $2\Delta_{0}/T_{sc}$ and $2\widetilde{\Delta}_{0}/T_{sc}$, which is $3.53$ for the BCS superconductivity (for FL-SC there is no difference between $\Delta_{0}$ and $\widetilde{\Delta}_{0}$ as discussed below). This is shown in Figs. \ref{fig:gap} (c) and (d). 
We find that in our case, this ratio approaches the BCS value for smaller $\theta$ (FL normal state) and weaker on-site interaction. For SC emerging from NFL normal state this ratio deviates strongly from the BCS value. The value of this ratio first increases with $\theta$ as long as the transition is continuous, and then tends to decrease as the transition changes its nature to first order. 
This trend follows from the observation that the transition temperature increases very sharply for large values of $\theta$ and $R/|U|$ compared to the much gradual increase in $\Delta_{0}$. 
In the FL case (smaller $\theta$), both $\Delta_{0}$ and $T_{sc}$ are suppressed exponentially as a function of $R/|U|$ such that their ratio is a constant. However, in the NFL case (larger $\theta$) this is not true anymore. Both $\Delta_{0}$ and $T_{sc}$ appear to decrease with different power-laws with respect to $R/|U|$, and in particular for larger values of $\theta$ the transition temperature, $T_{sc}$, saturates quickly for large $\theta$. This is shown in Fig. \ref{fig:R} in the Appendix. 

\begin{figure}
    \centering
    \includegraphics[width=\textwidth]{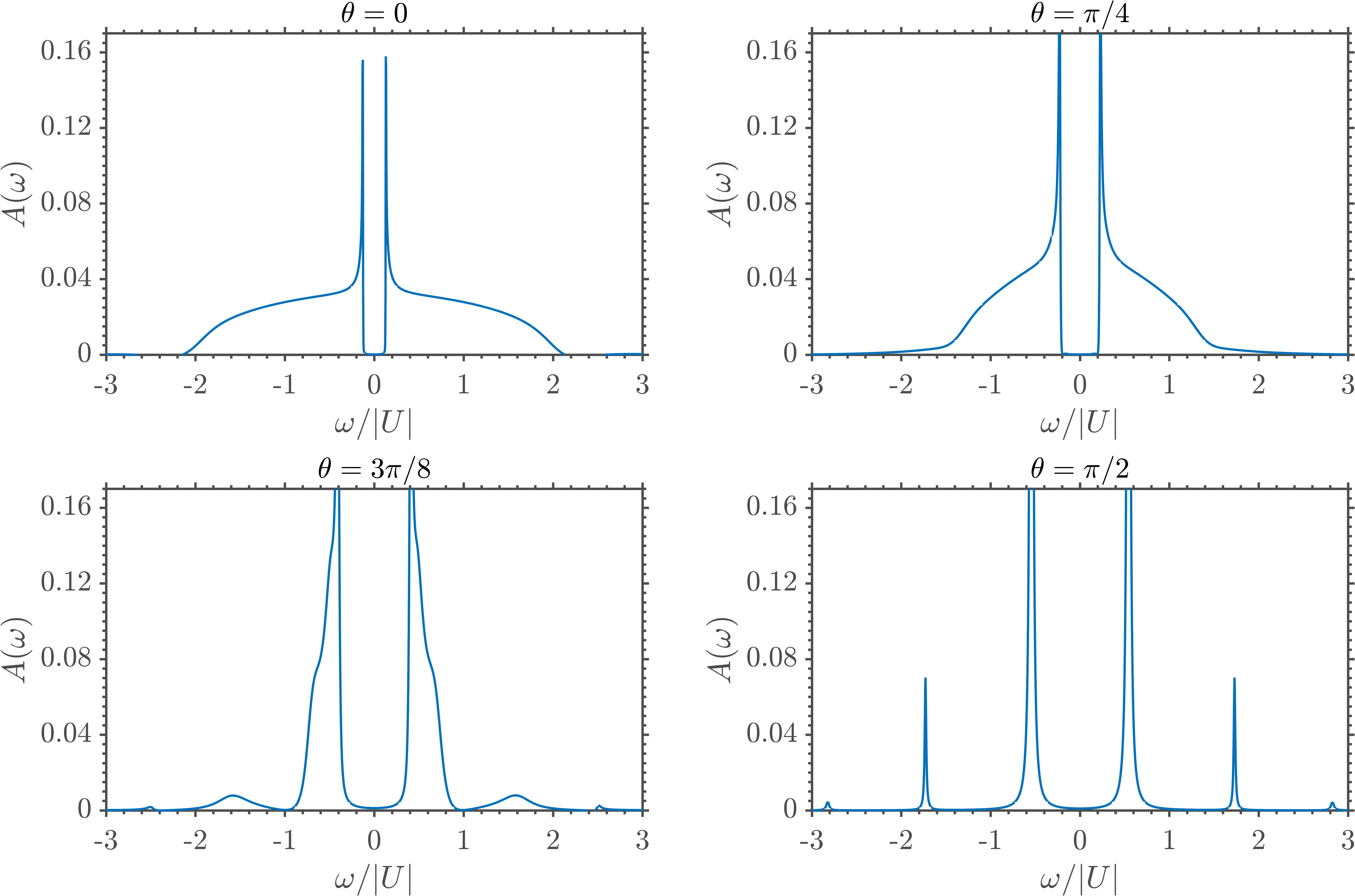}\\
     \caption{ The spectral functions in the superconducting phase at $R=|U|$ for different values of $\theta$. }
    \label{fig:spec_sc}
\end{figure}
\begin{figure}
    \centering
    \subfloat[]{\includegraphics[width=0.45\textwidth]{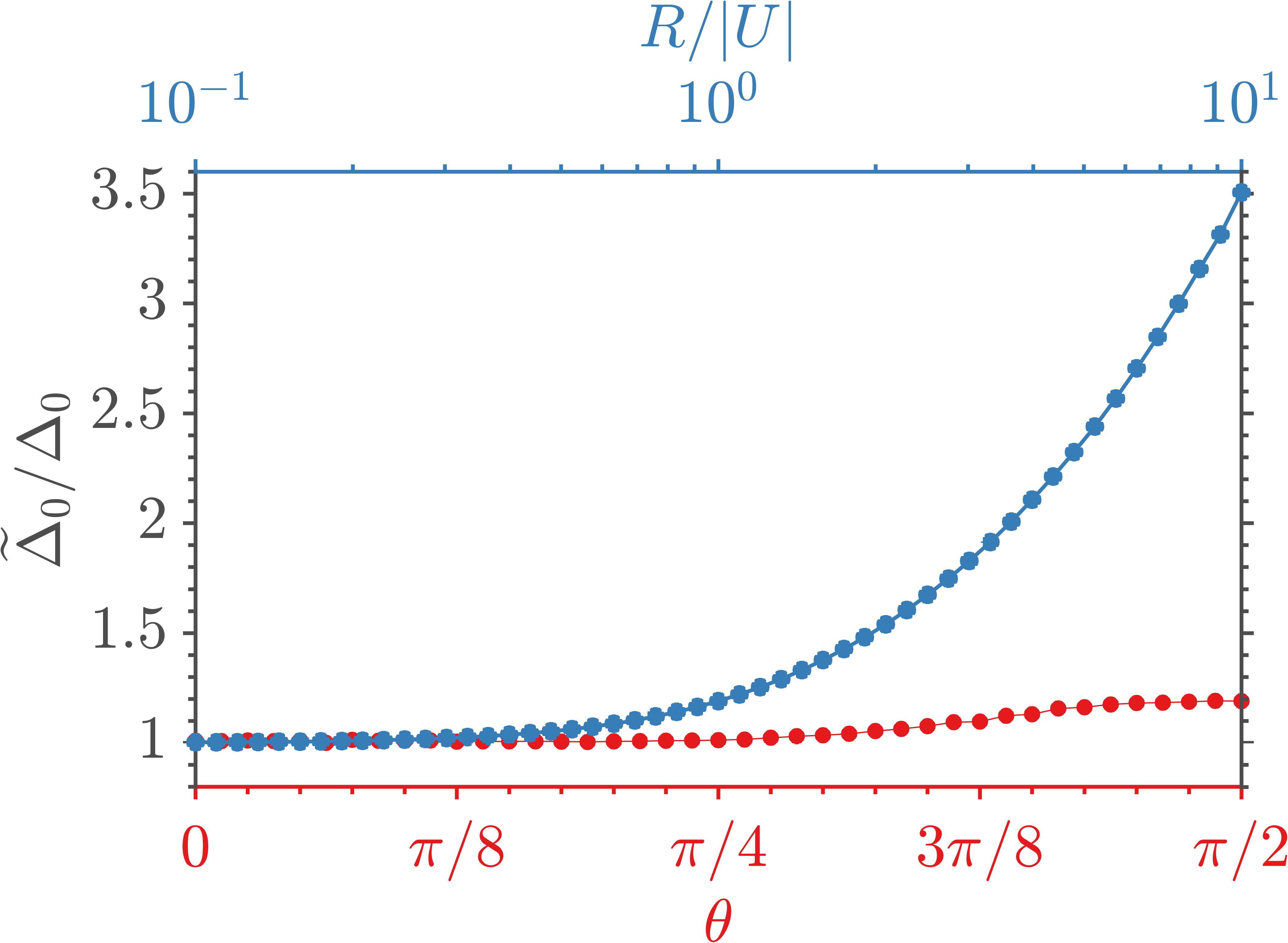}}~~~~
    \subfloat[]{\includegraphics[width=0.45\textwidth]{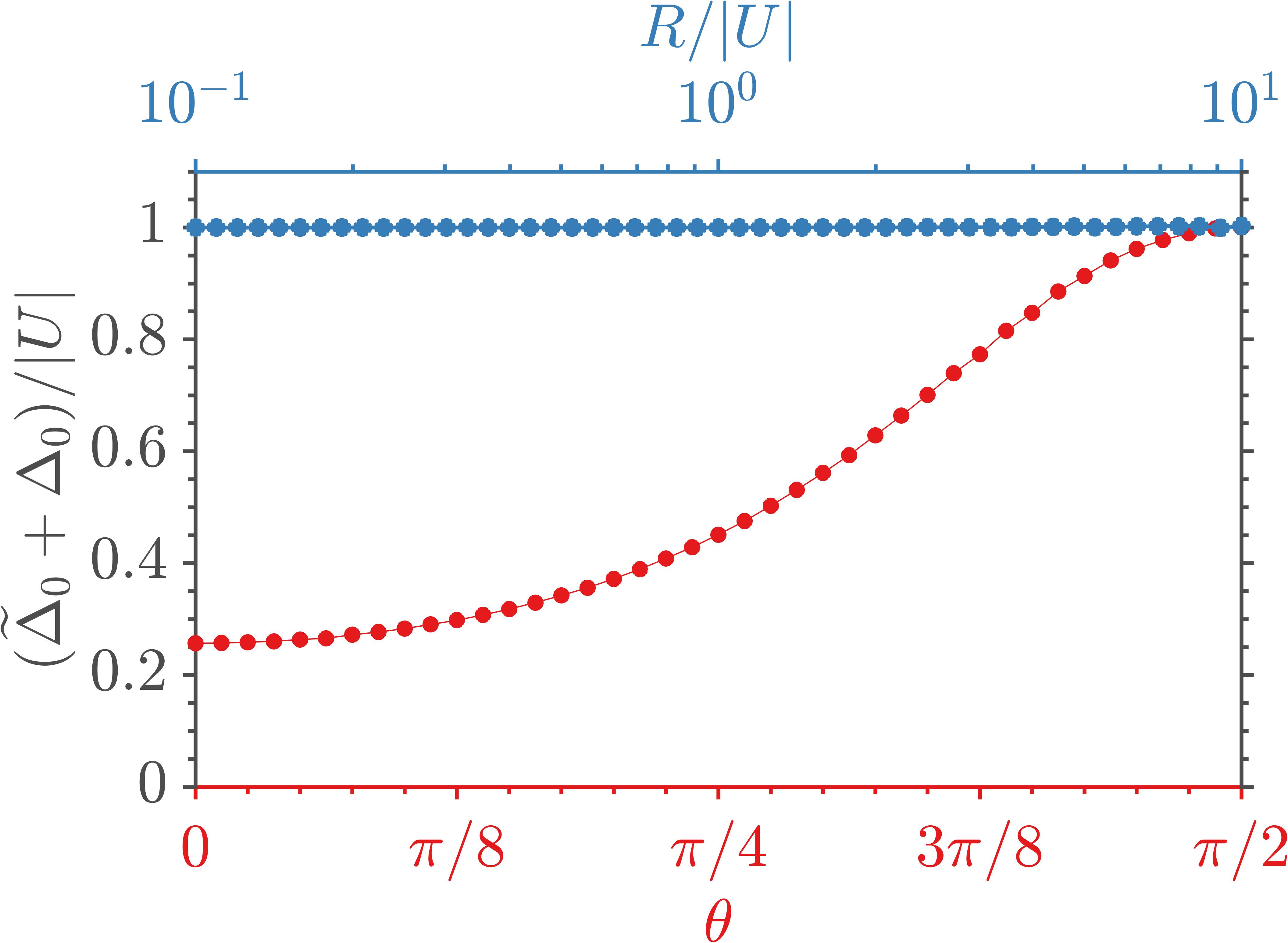}}
    \caption{ (a) In the limit of zero temperature, the ratio of the SC gap, $\widetilde{\Delta}_{0}$, (as obtained from the spectral function) and the SC order parameter, $\Delta_{0}$, as a function of $R/|U|$ at $\theta=\pi/2$ (blue) and as a function of $\theta$ at $R/|U|=1$ (red) is shown. The two quantities are in general different away from the FL limit and for small on-site interaction the deviation between the two quantities is strongest.
    (b) The sum $\widetilde{\Delta}_{0}$ and $\Delta_0$ as a function of $R/|U|$ at $\theta=\pi/2$ (blue) and as a function of $\theta$ at $R/|U|=1$ (red) is shown. The sum is a constant for $\theta=\pi/2$. However, this is not the case for other values of $\theta$. 
    }
    \label{fig:gapratio}
\end{figure}

We have also computed the spectral function for the SC phase. This is shown in Fig. \ref{fig:spec_sc}. As expected, we clearly see the SC gap in the spectral function. For $\theta=0$ (FL normal state) we see the expected square-root divergence near $\om=\Delta$. The form of this divergence seems to be modified for $\theta$ away from zero. In particular, for $\theta=\pi/2$ (SYK-NFL normal state) we see very narrow peaks. We also note that the SC gap ($\widetilde{\Delta}$) observed in the spectral function may not be the same as SC order parameter $\Delta$ calculated above, as is shown in Fig. \ref{fig:gap} (a) and (b). The two quantities are same for SC emerging from FL (smaller $\theta$), but may deviate from each other for the SC emerging from a NFL phase (larger $\theta$). In particular, the deviation between $\Delta$ and $\widetilde{\Delta}$ is strongest for larger $\theta$ and larger values of $R/|U|$ (where the transition is of first order). In Fig. \ref{fig:gapratio} (a), we show the variation of the ratio of these two quantities in the limit of zero temperature, i.e. $\widetilde{\Delta}_{0}/\Delta_{0}$, with respect to $\theta$ and $R/|U|$.  
We do not have an analytic expression for the gap in the spectral function, $\widetilde{\Delta}$.
But numerically we find that $\Delta_{0}+\widetilde{\Delta}_{0} \approx |U|$ at $\theta=\pi/2$, independent of the ratio $R/|U|$. This relation does not hold for other values of $\theta$. This is shown in Fig. \ref{fig:gapratio} (b). 

A noticeable new feature for SC emerging from NFL (larger values of $\theta$) is the presence of peaks at higher energies compared to the SC gap (see Fig. \ref{fig:spec_sc} (c) and (d)). In the limit of $T\to 0$ the first higher-order peak appears at $\sim 3\widetilde{\Delta}$. A dominant all-to-all exchange interaction (large $\theta$) means strongly interacting Cooper pairs, which may be the reason for these additional peaks. For smaller values of $\theta$ the Cooper pairs are weakly interacting. Note that such high energy features in the spectral function have also been reported for SYK-like electron-phonon model for SC \cite{Esterlis2019}. 

\begin{figure}
    \centering
    \includegraphics[width=\textwidth]{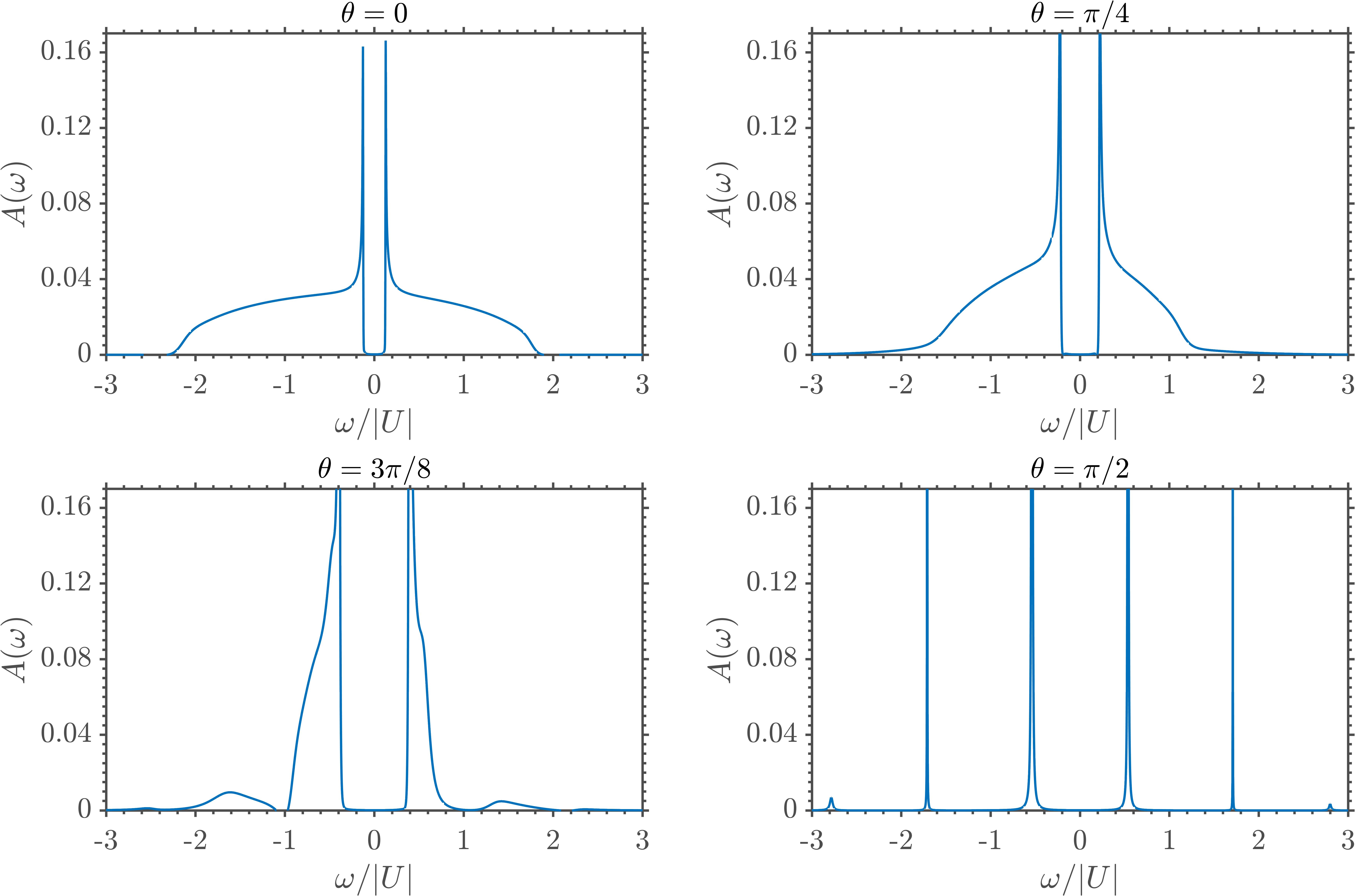}
    \caption{The spectral functions in the superconducting phase at $R = |U |$, $\mu/|U|=0.2$ for different values of $\theta$.}
    \label{fig:doped}
\end{figure}

We have so far focused on the particle-hole symmetric point, $\mu=0$, for clarity. However, it is straightforward to also do the same analysis for a non-zero chemical potential. The results are qualitatively the same as discussed above. The main difference seen is the particle-hole asymmetric distribution of the spectral weights at positive and negative frequencies, as shown in Fig. \ref{fig:doped}. The gap is however symmetric around $\omega=0$. In the remainder of the paper we again focus only on $\mu=0$.

\begin{figure}
\centering
\subfloat[]{\includegraphics[width=0.45\textwidth]{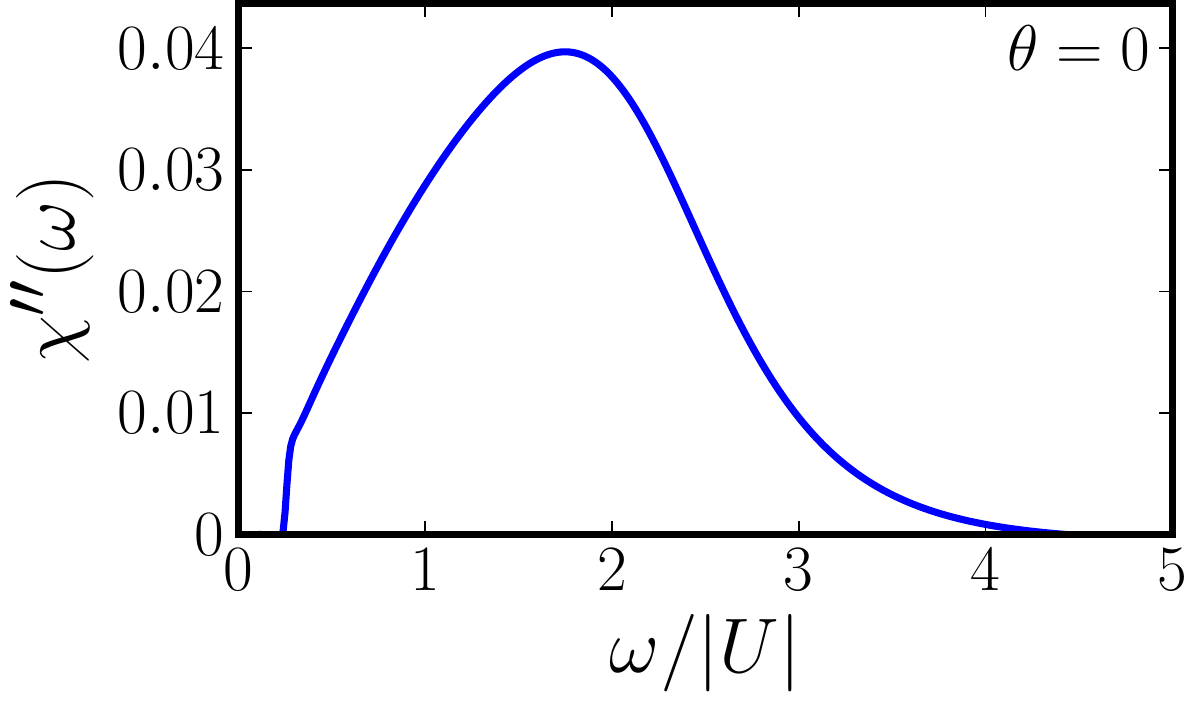}} ~~~~ 
\subfloat[]{\includegraphics[width=0.45\textwidth]{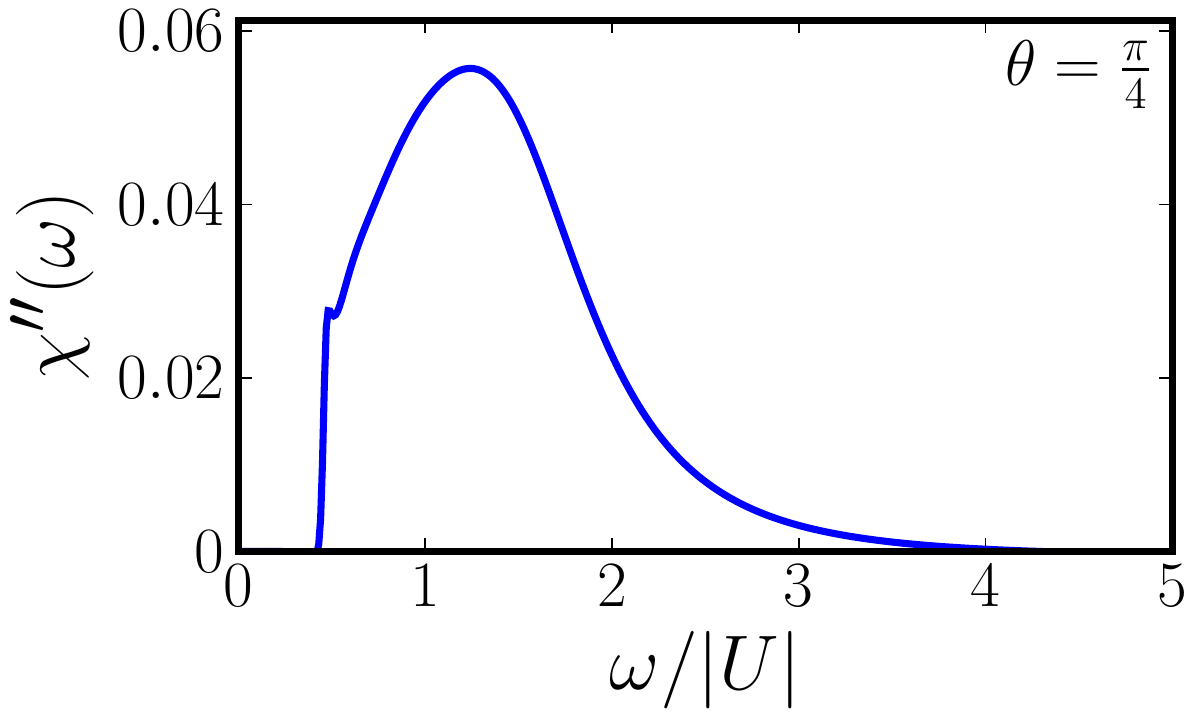}} \\
~~~\subfloat[]{\includegraphics[width=0.45\textwidth]{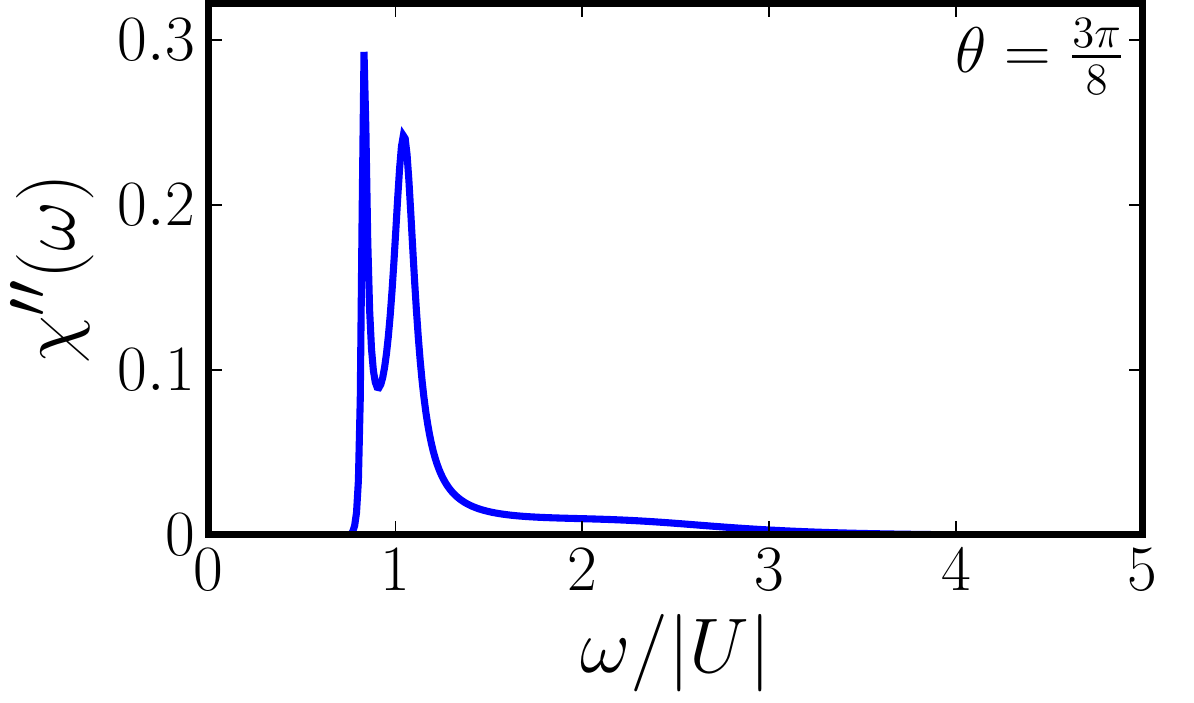}} ~~~~ 
\subfloat[]{\includegraphics[width=0.45\textwidth]{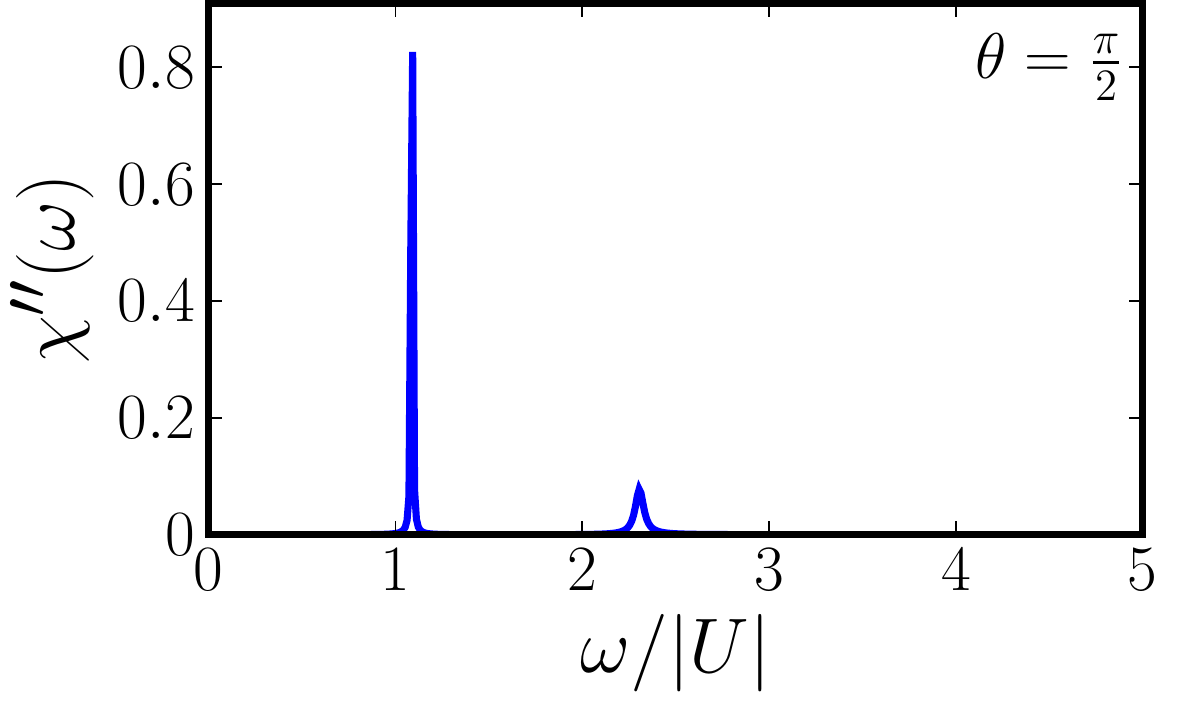}}
\caption{Plot of imaginary part of spin correlation $\chi''(\omega)$ in the SC phase as a function of real frequency for different values of $\theta$ at $R=|U|$ and $T=0.01$.}
\label{fig:chi_sc}
\end{figure}

We further also compute the spin correlation in the SC phase. In Fig. \ref{fig:chi_sc} we plot $\chi''(\omega)$ for different values of $\theta$ in the SC phase. The features essentially follow from what was discussed for the electron spectral function earlier. The high-energy peak present in the electron spectral function at larger values of $\theta$ is also seen in $\chi''(\omega)$.

\begin{figure}
\centering
\subfloat[]{\includegraphics[width=0.45\textwidth]{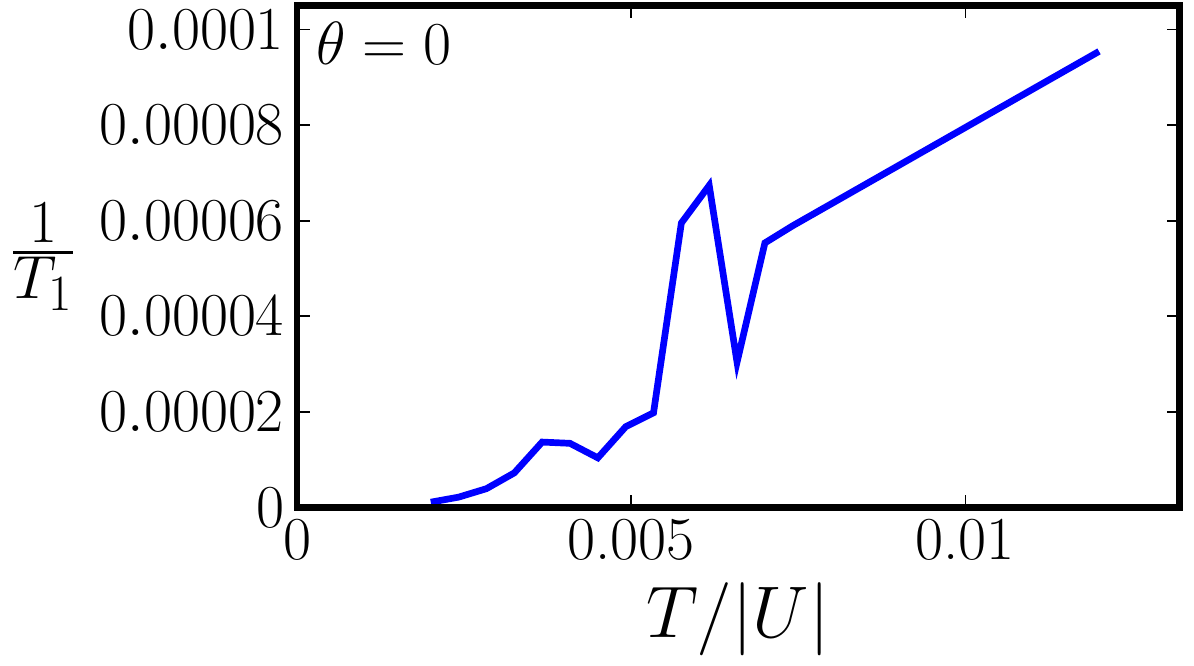}} ~~~
\subfloat[]{\includegraphics[width=0.45\textwidth]{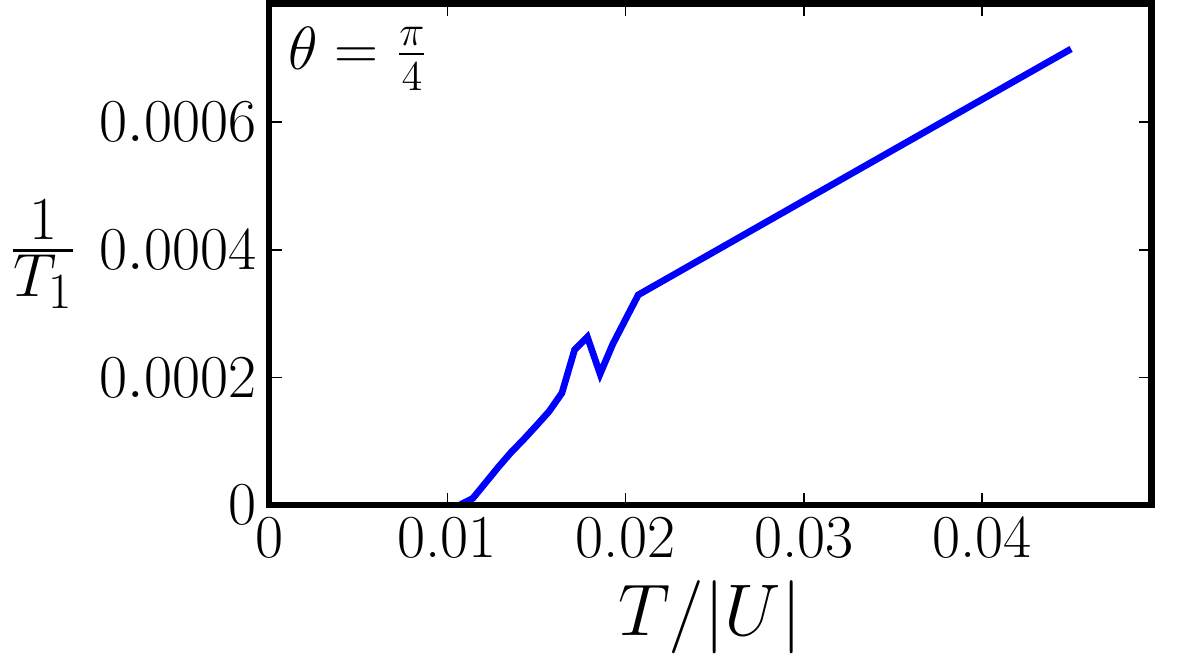}} \\
~~~~~\subfloat[]{\includegraphics[width=0.45\textwidth]{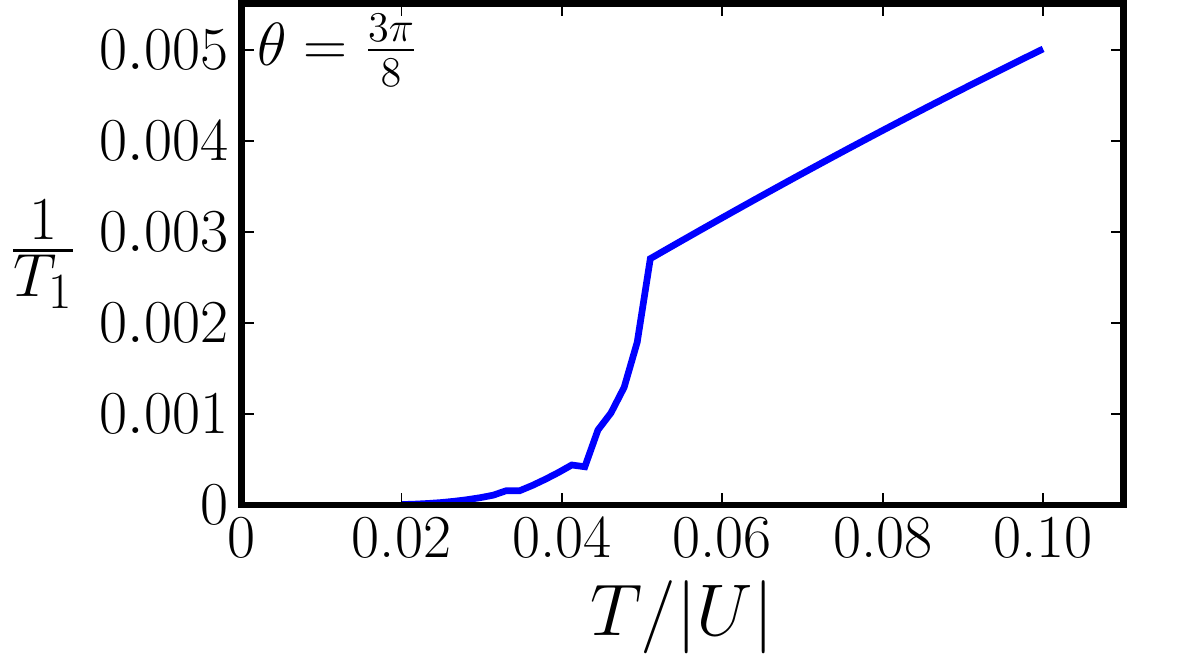}} ~~ 
\subfloat[]{\includegraphics[width=0.45\textwidth]{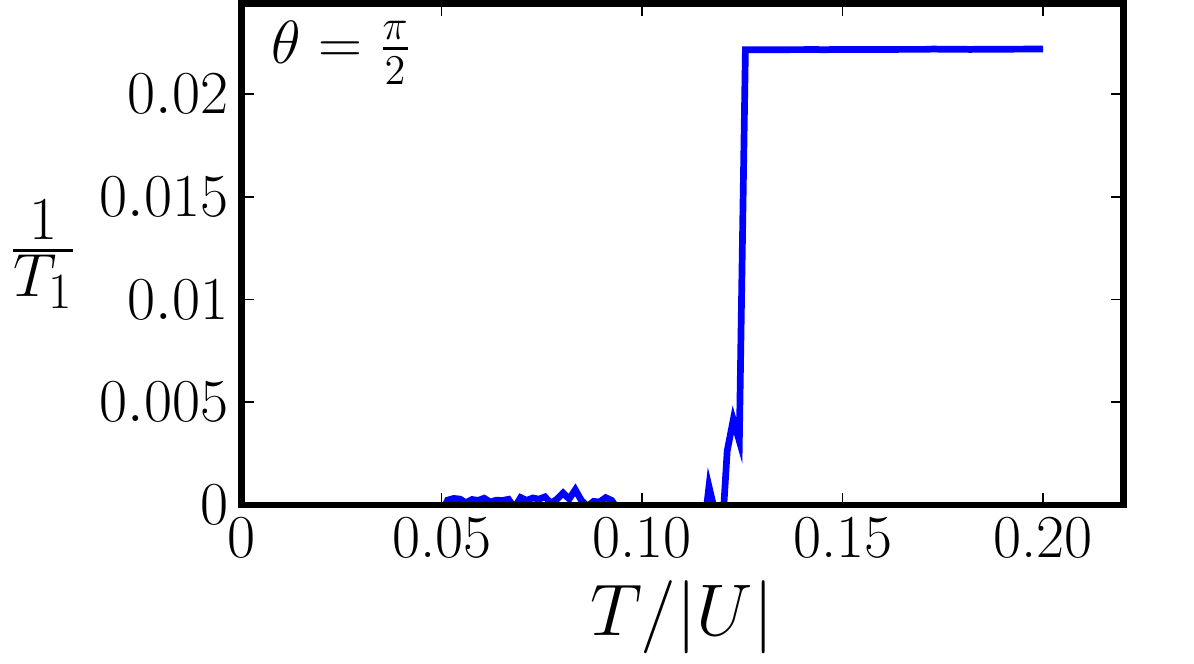}}
\caption{Temperature dependence of the NMR relaxation rate, $1/T_{1}$, for different values of $\theta$ at $R=2|U|$. Note that for smaller values of $\theta$, where the normal state is FL-like, there is a `Hebel-Slichter' peak around the SC transition temperature. This can be seen in (a) around $T/|U| \sim 0.006$ and in (b) around $T/|U| \sim 0.018$, which have FL normal state. Note that the peak height diminishes as we increase $\theta$ and go closer to NFL case. The `Hebel-Slichter' peak is absent (and replaced by a {\em kink}) in case of NFL normal state, shown in (c) and (d). 
}
\label{fig:nmr}
\end{figure}

Using $\chi''(\omega)$ we can also evaluate the temperature dependence of the NMR relaxation rate, $1/T_{1}$, which we show in Fig. \ref{fig:nmr}. The NMR relaxation rate is given by the following relation \cite{PG98}:
\begin{equation}
\label{eq:nmr}
\frac{1}{T_{1}} = T \left. \frac{\chi''(\omega)}{\omega} \right|_{\omega = 0} \,.
\end{equation}
Recall that for the standard BCS superconductor one expects a peak (often referred to as the `Hebel-Slichter' peak) around the critical temperature as a consequence of the square-root divergence in the spectral function \cite{HS}.  However, one of the signatures of the unconventional superconductivity is the absence of `Hebel-Slichter' peak, for instance, as observed in cuprates \cite{Takigawa} and Fe-based superconductors \cite{Fukazawa}. We find that for a fixed $R/|U|$ when $\theta \lesssim \theta_{coh}$ there is a well distinguished `Hebel-Slichter' peak whose strength diminishes with increasing $\theta$ (see Fig. \ref{fig:nmr} (a) and (b)). After the crossover into the NFL regime for larger $\theta$ the peak is absent and there is only a {\em kink} around the critical temperature (see Fig. \ref{fig:nmr} (c) and (d)). This is another distinguishing feature between the FL and NFL case. We also see that the relaxation rate is higher for the NFL case compared to the FL case. In the normal state this trend easily follows from the fact that the critical temperature is much higher in the NFL case. Also note that the height of the `Hebel-Slichter' peak present for smaller $\theta$ is roughly inversely proportional to $R/|U|$, i.e., for smaller Hubbard interaction the peak is smaller.

\section{Discussion}
\label{sec:dis}

We have investigated the emergence of SC in a SYK-like model of interacting electrons, Eq.~(\ref{eq:Ham}). The model is solved in the large-$M$ limit, where we generalize the spin symmetry from SU$(2)$ to SU$(M)$. The solution of the large-$M$ saddle-point equations can be viewed as a dynamical mean-field solution. We have shown the contrast between the emergence of SC from a NFL as opposed to a FL normal state. Several distinguishing features are found for SC emerging from a NFL and we summarize below the salient features of our work.

\begin{itemize}
\item Even in the presence of all-to-all and random exchange interaction and Cooper-pair hopping, we show that BCS-type superconducting instability is present, thus ensuring SC ground state at zero temperature for any infinitesimal attractive Hubbard interaction.

\item The SC transition temperature, $T_{sc}$, is shown to be strongly enhanced for NFL normal state as compared to a FL normal state. This is an important highlight of our results. This is understood physically by realizing that the most dominant mechanism to break Cooper pairs is single-particle hopping. However for the NFL case the Cooper pairs are strongly interacting, and single-particle hopping is sub-dominant, thus leading to a higher $T_{sc}$. This also renders the transition in case of NFL to be first order for weaker Hubbard interaction. 

\item While for the FL (BCS-like) case both $T_{sc}$ and $\Delta$ are exponentially suppressed with respect to $R/|U|$, we show that for NFL case they decay with different power-laws. Consequently, the ratio $2\Delta/T_{sc}$ strongly deviates from the BCS value for SC arising from NFL.

\item We have presented a detailed study of the local electron spectral function in the SC as well as the normal states (FL and NFL). This is an observable in photoemission experiments like ARPES. We discuss how the SC gap closes upon approaching $T_{sc}$. In the case of a FL normal state the transition is continuous and BCS like, and the spectral function in SC phase features the well-known square-root divergence at $\omega=\Delta$. We show that this is not the case when the normal state is a NFL. 

\item We show that for SC emerging from a NFL there is a distinct new feature in the local electron spectral function -- peaks at higher energy at $\omega \sim 3\Delta$. This is a consequence of strong interactions between Cooper pairs (which is absent in case of FL normal state). We believe that this is a generic feature of SC emerging from a NFL, and could be a relevant observation in many materials. How generic and model independent is this feature is an interesting open question for future. 

\item In the normal state, as a function of the parameter $\theta$, there is a crossover between FL and NFL phase for a fixed temperature, which we characterize using the effective local spin correlation exponent, $\eta_{S}$. The exponent deviates from FL value  for $\theta \gtrsim \theta_{coh}$. We hope that our work motivates the observation of this exponent in neutron scattering experiments.

\item We also note that NFL phase, i.e., the normal state for $\theta > \theta_{coh}$ has a linear-in-temperature resistivity. Thus SC emerging from this state may have some relevance to the situation in correlated systems. 

\item We have evaluated the local dynamic structure factor, $\chi''(\omega)$, an observable in neutron scattering experiments. In the SC phase emerging from NFL $\chi''(\omega)$ shows distinct peaks at high energies akin to that discussed for the spectral function. 

\item Further we have also calculated the NMR relaxation rate, $1/T_{1}$, as a function of temperature. Here we show that for FL normal state there is a `Hebel-Slichter' peak near $T_{sc}$, which is a hallmark of BCS SC. However for NFL case this peak disappears and the transition temperature is marked by a {\em kink}. Such observations have been reported in experiments on unconventional SC in cuprates and pnictides. Our work clearly shows the mechanism for the disappearance of the `Hebel-Slichter' peak in the case of a NFL normal state. This may be of general relevance to the NMR experiments in unconventional SC materials.
\end{itemize}

We believe that our work will further motivate and provide a pathway to investigate SC emerging from NFL. 
Our work also motivates numerical investigation of the model in Eq. (\ref{eq:Ham}) at $M=2$ to further elucidate the SC-NFL phase transition. We hope that our work may also provide a good starting point for constructing more realistic lattice models. 

While this work was being completed, we learnt of the study Ref.~\cite{Chudnovskiy2022} of essentially the same model, but with a focus on the finite $N$ behavior.


\section*{Acknowledgements}

We thank G. Tarnopolsky for valuable discussions. This research was supported by the National Science Foundation under Grant No. DMR- 2002850. This work was also supported by the Simons Collaboration on Ultra-Quantum Matter, which is a grant from the Simons Foundation (651440, S.S.).
D.G.J. acknowledges support from the Leopoldina fellowship by the German National Academy of Sciences through grant no. LPDS 2020-01.


\appendix

\section{Numerical analytic continuation}
\label{sec:nac}

We also perform numerical analytic continuation to real frequency. In general, performing analytic continuation is an ill-posed problem if the function on the imaginary axis is known only at a finite number of points. There are several techniques to do analytic continuation. But for simplicity we use the Pade approximation method. This technique parametrizes the function on imaginary axis as a ratio of two polynomials or by terminating a continued fraction. There are several ways for implementing Pade approximation. We adopt the simple strategy outlined in Ref. \cite{Vidberg_Serene} of evaluating the coefficients of the two polynomials recursively, which is based on Thiele's reciprocal difference method. Details of the algorithm can be found in the Appendix of Ref. \cite{Vidberg_Serene}. Briefly, we first solve the saddle-point equations on the imaginary-frequency axis to obtain the required Green's function, say $G(\iw)$, at non-negative Matsubara frequencies. The number of Matsubara frequencies used in our calculation is $10^5$. Then we evaluate the required polynomials, $A_{n}(z)$ and $B_{n}(z)$, to approximate the imaginary-frequency function, $G(z) = A_{n}(z)/B_{n}(z)$. The accuracy of these polynomials depends on the number of Pade points, $n$, and in our calculation we find that $n=200$ points are sufficient to obtain accurate results. We have checked our results by increasing or decreasing $n$ and it does not result in any significant improvement. The resulting ratio of polynomials then corresponds to the retarded Green's function on real-frequency axis, once we identify $z=\om+i0^{+}$. Imaginary part of this function then gives the spectral function.

\section{Effective spin exponent}
\label{sec:etaS}

\begin{figure}
\centering
\subfloat[]{\includegraphics[width=0.45\textwidth]{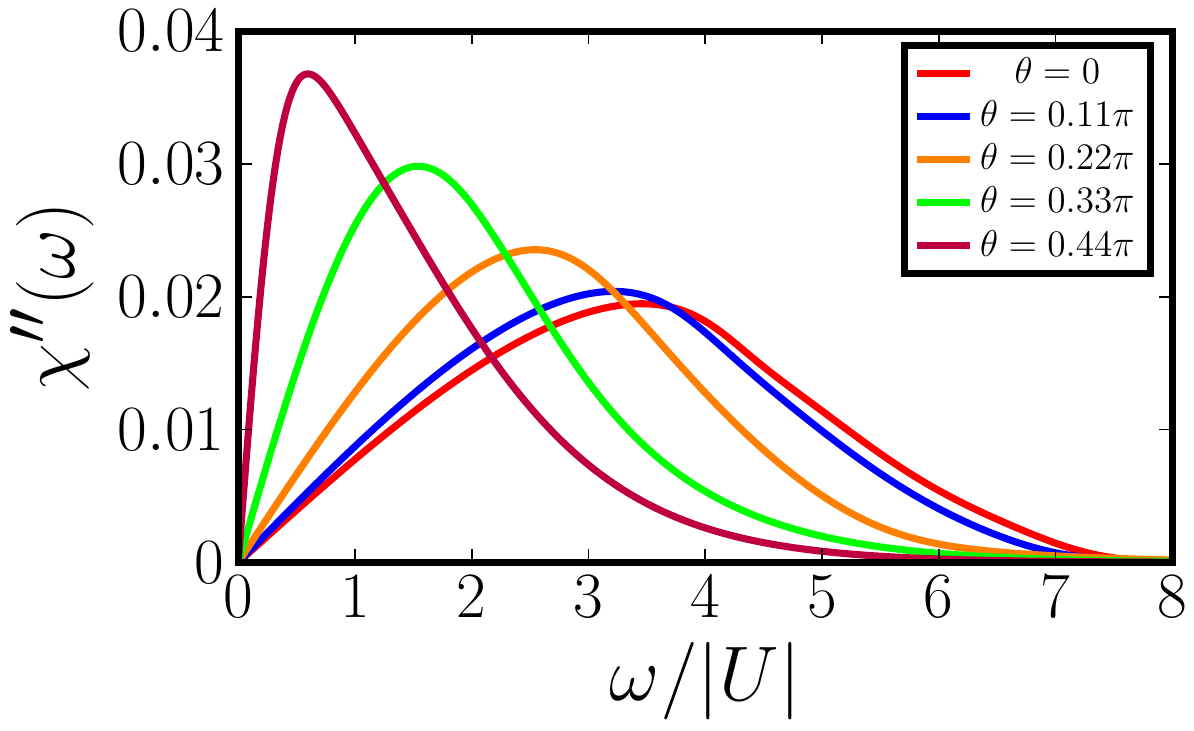}} ~~~~
\subfloat[]{\includegraphics[width=0.45\textwidth]{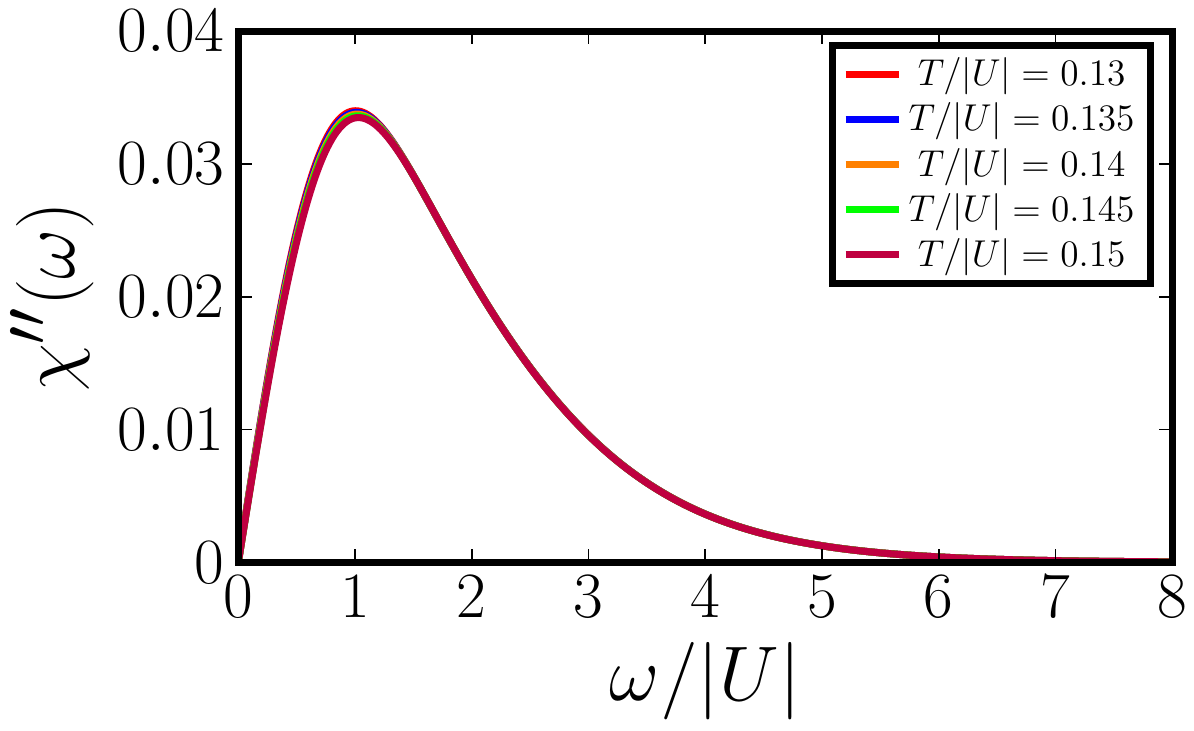}}
\caption{(a) Plot of $\chi''(\omega)$ for different values of $\theta$ at a fixed temperature $T/|U|=0.13$ in the normal state. 
(b) Plot of $\chi''(\omega)$ for different values of temperature at a fixed value of $\theta = 0.388\pi$. In both the plots, $R/|U| =2$.}
\label{fig:chi}
\end{figure}

In this appendix we discuss the evaluation of the effective spin exponent ($\eta_{s}$) in the normal state. To start with, we first evaluate the spin correlation, $\chi(\tau) = \langle \vec{S}(\tau) \cdot \vec{S}(0) \rangle \sim -G(\tau) G(-\tau)$, which is straightforward to obtain from the imaginary frequency numerics. We then Fourier transform to obtain $\chi(\iw)$, and then perform numerical analytic continuation to obtain $\chi(\omega)$ whose imaginary part is the dynamical susceptibility, $\chi''(\omega)$. This is shown in Fig. \ref{fig:chi} for the normal state and in Fig. \ref{fig:chi_sc} in the SC phase. 

At temperature above the SC transition temperature, the normal state solution is one of the SYK-type conformal solutions at low energy ($\omega \ll \tilde{J}$). For such a solution the spin susceptibility follows the scaling relation \cite{Sachdev94_afm, Parcollet98,PG98},
\begin{equation} 
\chi''(\omega) \sim T^{\eta_{s} -1} \Phi_{\eta_{s}} \left( \frac{\hbar \omega}{k_{B} T} \right) \,,
\end{equation}
where 
\begin{equation}
\Phi_{\eta_{s}} (y) = \sinh \left(\frac{y}{2} \right) \left|\Gamma \left(\frac{\eta_{s}}{2} + i\frac{y}{2\pi} \right) \right|^{2} \,.
\end{equation}
For $\hbar \omega \ll k_{B} T$ we have,
\begin{equation}
\label{eq:chi2}
\chi''(\omega) \sim \omega ~ T^{\eta_{s} -2} \,,
\end{equation}
while in the limit of $\hbar \omega \gg k_{B} T$, the result is similar to the zero temperature form, $\chi''(\omega) \sim \sgn(\omega) |\omega|^{\eta_{s}-1}$. 

\begin{figure}
\centering
\includegraphics[width=0.55\textwidth]{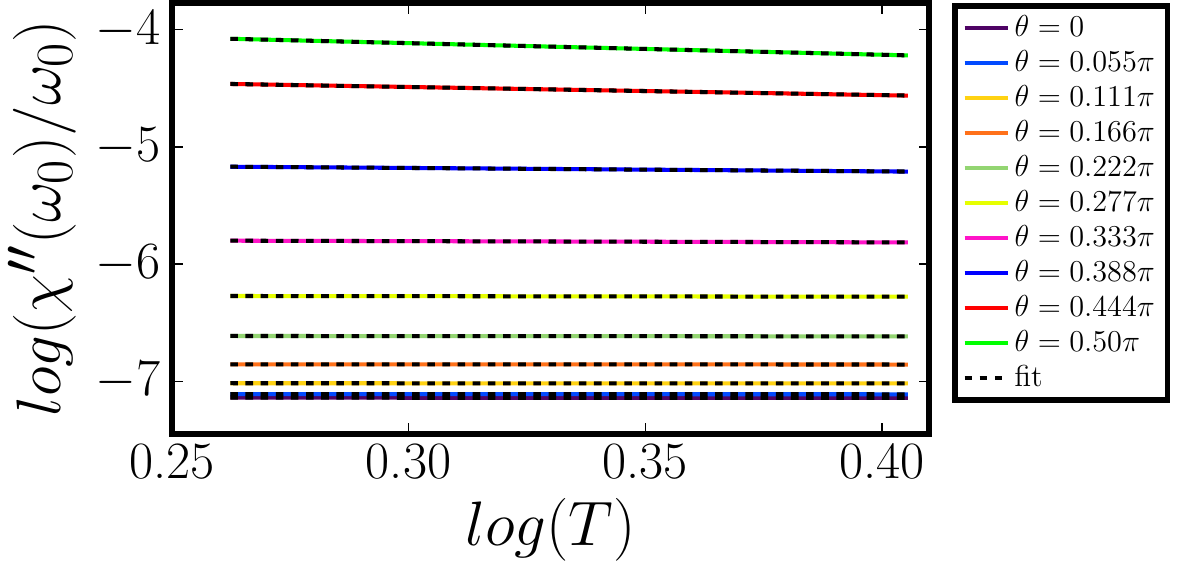}
\caption{Plot of $\log(\chi''(\omega_{0})/\omega_{0})$ versus $\log(T)$ at $\omega_{0} =0.2$, $R/|U|=2$, and in the temperature range between $T/|U| = 0.13$ and $T/|U| = 0.15$, for different values of $\theta$. The slope of the linear fit (black dashed lines) gives $\eta_{s}-2$, from Eq. (\ref{eq:chi2}). This is used to plot the curve of $\eta_{s}$ as a function of $\theta$ in Fig. \ref{fig:Aw_ns} (b).}
\label{fig:fit_chi}
\end{figure}

We can thus use Eq. (\ref{eq:chi2}) to extract the effective spin exponent ($\eta_{s}$) from the slope of plot of $\log(\chi''(\omega_{0})/\omega_{0})$ versus $\log(T)$, where $\omega_{0}$ is a fixed small frequency. In Fig. \ref{fig:fit_chi} we present the data for such a procedure for $R/|U|=2$ in the temperature range of $T/|U| = 0.13$ and $T/|U| = 0.15$, with $\omega_{0}=0.2$. We have also checked our results for two other small frequency points, and the results are unchanged. The resulting $\eta_{s}$ as a function of $\theta$ is plotted in Fig. \ref{fig:Aw_ns} (b). Similar procedure can be done at other values of $R/|U|$. This works well for larger values of $R/|u|$. At smaller values of $R/|U|$ the transition temperature is relatively high, where our numerical analytic continuation is not very reliable, and so extracting $\eta_{s}$ there is difficult. 

\begin{figure}
\centering
\subfloat[]{\includegraphics[width=0.45\textwidth]{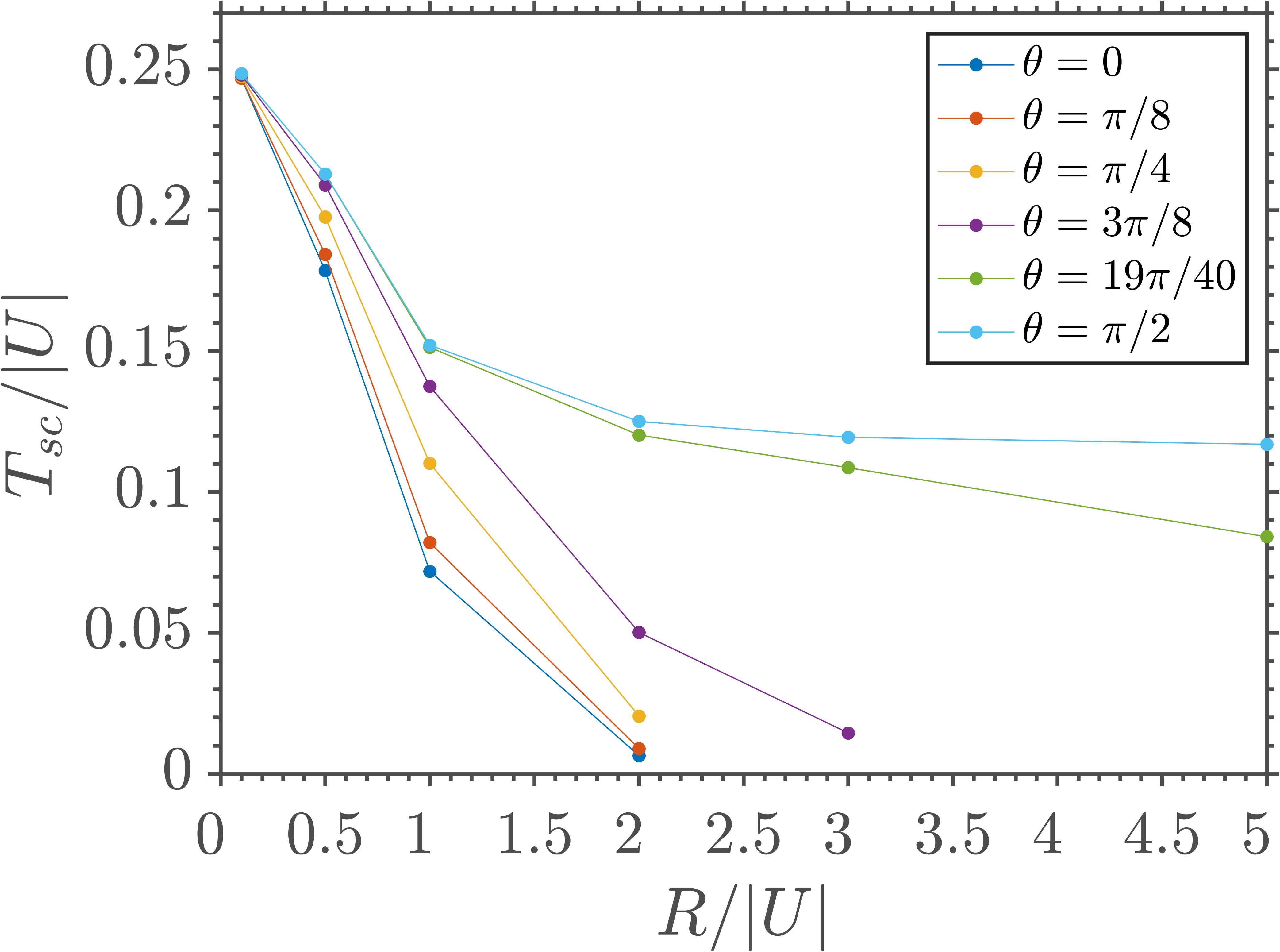}} ~~~~
\subfloat[]{\includegraphics[width=0.45\textwidth]{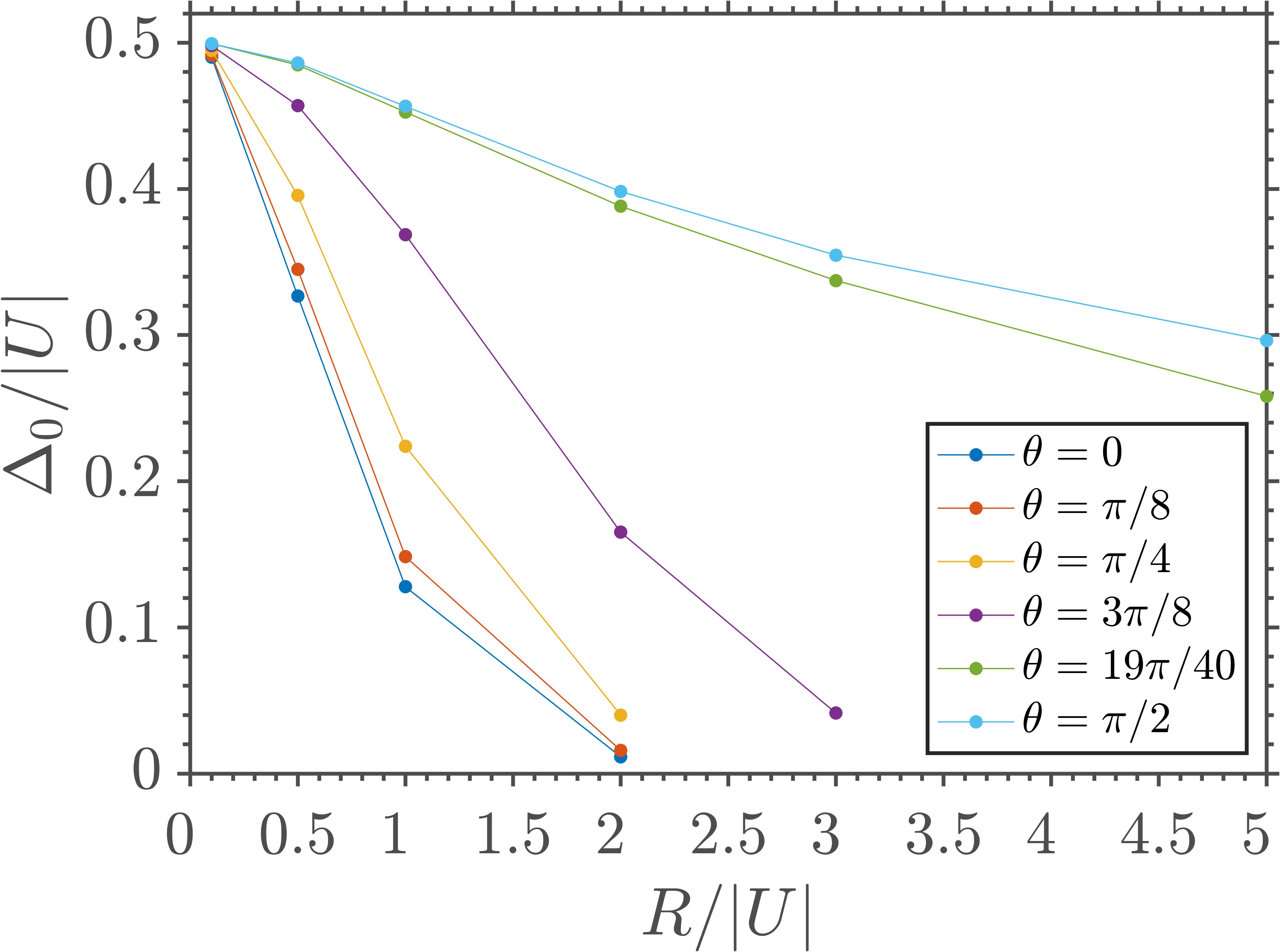}} 
\caption{Plots of transition temperature $T_{sc}$, the SC order parameter $\Delta_0$ and the SC gap $\widetilde{\Delta}_0$ versus $R$ for different values of $\theta$. Note that for smaller values of $\theta$ when the normal state is FL-like, there is an exponential decay with respect to $R/|U|$. In contrast, in the case of NFL normal state these are replaced by different power-laws. 
}
\label{fig:R}
\end{figure}

\bibliography{scmt}

\end{document}